\documentclass[a4paper,10pt,twoside]{cpc-hepnp}
\usepackage{CJK,upgreek,fancyhdr}
\usepackage{multicol}
\usepackage{graphicx}
\usepackage{booktabs}
\usepackage{amssymb,bm,mathrsfs,bbm,amscd}
\usepackage[tbtags]{amsmath}
\usepackage{lastpage}
\usepackage{color}
\usepackage{slashed}
\usepackage{graphicx}
\usepackage{upgreek}
\usepackage{url}
\usepackage{lineno}
\usepackage{multirow}
\usepackage[english]{babel}
\addto{\captionsenglish}{%

}

\begin{document}
\begin{CJK*}{GB}{gsbn}


\title{A mass reconstruction technique for a heavy resonance  decaying to $\uptau^+\uptau^-$\thanks{Supported by the General Financial Grant from the China Postdoctoral Science Foundation (Grant No. 2015M581062).}}

\author{Li-Gang Xia  \email{xialigang@tsinghua.edu.cn}}
\maketitle
\address{Department of Physics, Tsinghua University, Beijing 100084, People's Republic of China}

\begin{abstract}
For a resonance decaying to $\uptau^+\uptau^-$, it is difficult to reconstruct its mass accurately because of the presence of neutrinos in the decay products of the $\uptau$ leptons. If the resonance is heavy enough, we show that its mass can be well determined by the momentum component of the $\uptau$ decay products perpendicular to the velocity of the $\uptau$ lepton, $p_{\perp}$, and the mass of the visible/invisible decay products, $m_{vis/inv}$, for $\uptau$ decaying to hadrons/leptons. By sampling all kinematically allowed values of $p_{\perp}$ and $m_{vis/inv}$ according to their joint probability distributions determined by the MC simulations, the mass of the mother resonance is assumed to lie at the position with the maximal probability. Since $p_{\perp}$ and $m_{vis/inv}$ are invariant under the boost in the $\uptau$ lepton direction, the joint probability distributions are independent upon $\uptau$'s origin. Thus this technique is able to determine the mass of an unknown resonance with no efficiency loss. It is tested using the MC simulations of the physics processes $\rm pp \to Z/h(125)/h(750)+X\to \uptau\uptau+X$ at 13~TeV. The ratio of the full width at half maximum and the peak value of the reconstructed mass distribution is found to be 20\%-40\% using the information of missing transverse energy.

\end{abstract}
\begin{keyword}
Mass reconstruction; $\uptau$ lepton; Higgs boson
\end{keyword}
\begin{pacs}
29.85.Fj
\end{pacs}




\section{Introduction}
Following the discovery of the Standard Model (SM) Higgs boson~\cite{higgs_atlas, higgs_cms}, it is of more interest to search for new high-mass particles, such as additional Higgs bosons~\cite{hmhiggs_atlas1, hmhiggs_atlas2, hmhiggs_cms1, hmhiggs_cms2}. Recently, in the $ \upgamma\upgamma$ mass spectrum, a resonance around the mass $750$~GeV/$c^2$ is seen with a significance of about $3\sigma$~\cite{h750_atlas, h750_cms}. In the SM and many beyond-SM theories, the coupling of the Higgs boson and the fermion is proportional to the mass of the fermion. For a neutral Higgs boson, $h^0$, the branching fraction of $h^0 \to\uptau^+\uptau^-$ would be dominant compared to other leptons $\rm e$ and $\upmu$. Experimentally, the presence of neutrinos makes it challenging to reconstruct the invariant mass $M(\uptau\uptau)$ accurately~\cite{svfit,htt_atlas}. Accordingly, the performance of the mass reconstruction technique affects the sensitivity of probing new physics. 

The $\uptau$ lepton has two types of decay modes. One is decaying to a charged lepton plus two neutrinos, $\rm \uptau \to l+\bar{\upnu_l}+\upnu_\uptau, l=e,\upmu$ (denoted by $\uptau_l$), the other is decaying to hadron(s) plus one neutrino, $\uptau \to \text{hadrons}+\upnu_\uptau$ (denoted by $\uptau_h$).  Experimentally, the visible decay product is the charged lepton ($\rm e,\upmu$) or the hadron(s) while the invisible decay product is the neutrino(s).  For the charged leptons, the momentum can be measured well by the tracking system. For the hadronic decay products, they are reconstructed from the energy clusters in the electromagnetic and hadronic calorimeters using the anti-$k_t$ jet finding algorithm~\cite{anti_kt1, anti_kt2}. One of the characteristics is the presence of one or three charged tracks accompanied by possibly neutral hadrons, which lead to a collimated shower in the calorimeters with a few nearby tracks. The energy deposits in the calorimeters are used to reconstruct the four-momenta of the hadronic $\uptau$ candidates. Both the tracking and calorimeter information are combined to identify the hadronic $\uptau$ decays and suppress the misidentification rate of jets. 

For the $M(\uptau\uptau)$ reconstruction, an important observable is the missing transverse energy $\slashed{E}_{x/y}^{obs}$. They are defined as
\begin{equation}\label{eq:met_def}
\slashed{E}_{x/y}^{obs} \equiv -1 \times \sum_{\text{observed objects}} p_{x/y} \:,
\end{equation}
where the colliding axis is defined as the $z$ axis and the sum is performed over all observed objects, including electrons, muons, hadronic taus, jets, etc.. If only the neutrinos from the $\uptau$ decays contribute to the missing energy, $\slashed{E}_{x/y}^{obs}$ can then be used to improve the $M(\uptau\uptau)$ reconstruction, as will be shown below. Here we only use the transverse part of the missing energy. This is because in the hadron colliders, the interaction happens among the colliding partons. The partons in the high-boosting hadrons, such as proton, would have momenta collinear with the original momentum of the proton, and thus the initial transverse momenta are negligible~\cite{peskin}. It should be noted that the pileup effect would worsen the measurement of the missing transverse energy. For the CMS detector, the determination of the missing energy can be improved to have a resolution of 10-15~GeV~\cite{svfit} based on the data taken at 8~TeV using the particle-flow technique~\cite{met_cms}. For the ATLAS detector, a track-based soft term (TST)~\cite{TST_atlas} is adopted to mitigate the pileup effect and the missing transverse energy can be determined with a resolution of 10-20~GeV based on a small fraction of the data taken in 2015~\cite{met_atlas_2015}.

There are a few techniques of $M(\uptau\uptau)$ reconstruction which are commonly used in the hadron colliders. The study of these techniques can be found in Ref.~\cite{masscon_atlas, masscon_cms, MMC, svfit}. Here is a short summary. 
\begin{enumerate}
\item The Transverse Mass Method (TMM): $M(\uptau\uptau)=\sqrt{(P_{vis_1}+P_{vis_2}+P_{\slashed{E}})^2}$. Here $P_{vis_1}$ and $P_{vis_2}$ are the 4-momenta of the visible decay products of the $\uptau$ leptons; $P_{\slashed{E}} \equiv (\slashed{E}_x^{obs}, \slashed{E}_y^{obs}, 0, \sqrt{\slashed{E}_x^{obs2}+\slashed{E}_y^{obs2}})$. Furthermore, it is not able to consider the $z$ component of the momentum of the missing neutrinos. The resulting $M(\uptau\uptau)$ spectrum is thus biased and broadened.  
\item The Collinear Approximation Technique (CAT): It assumes that the neutrino(s) from each $\uptau$ decay is collinear with the visible $\uptau$ decay products, the charged leptons or the hadronical $\uptau$ jets. Thus the 4-momenta of the neutrinos can be fully determined. The $M(\uptau\uptau)$ is calculated in the equation below.
\begin{equation}\label{eq:mass_cat}
M(\uptau\uptau) = \sqrt{(P_{vis_1}+P_{vis_2})^2}/\sqrt{x_1x_2} 
\end{equation}
Here $x_1$ and $x_2$ are the momentum fractions carried by the visible $\uptau$ decay products. The explicit expressions are written below.
\begin{equation}\label{eq:cat}
x_{1/2} = \frac{p_{vis_{1/2}}}{p_{vis_{1/2}} + \slashed{E}_T(\cos(\phi_{vis_{1/2}}-\phi_{\slashed{E}})-\sin(\phi_{vis_{1/2}}-\phi_{\slashed{E}})\cot(\phi_{vis_{1/2}}-\phi_{vis_{2/1}}))}
\end{equation}
Here $p_{vis_{1/2}}$ is the magnitude of the momentum of the visible $\uptau$ decay products; $\slashed{E}_T \equiv \sqrt{\slashed{E}_x^2 + \slashed{E}_y^2}$; and $\phi_{vis_1/vis_2/\slashed{E}}$ are the azimuthal angles of the visible $\uptau$ decay products and the missing energy in the plane perpendicular to the beam line. This method gives a reasonable mass resolution, but it fails if the visible $\uptau$ decay products are outgoing back-to-back (thus $|\phi_{vis_1}-\phi_{vis_2}|\to\pi$ and $x_{1/2}\to 0$). It has an efficiency loss 30\%-60\%.     
\item The Missing Mass Calculator (MMC) Technique: The unknown quantities are the 4-momenta of the invisible neutrinos, denoted by $P_{inv_1}$ and $P_{inv_2}$. But there are only four constraints, written in Eq.~\ref{eq:mmc}. In Eq.~\ref{eq:mmc}, the first two constraints are from the information of missing energy and the last two constraints are due to the $\uptau$ lepton mass, denoted by $m_\uptau$. For each event, the method is to calculate all possible $M(\uptau\uptau)$ values satisfying the constraints by scanning the parameter space of the unknown variables. Each $M(\uptau\uptau)$ value carries a weight, which is the probability product $\mathcal{P}(\Delta R_1)\times \mathcal{P}(\Delta R_2)$. Here $\Delta R$ is the direction distance between the visible part and the invisible part for each $\uptau$ decay. It is defined as $\sqrt{(\eta_{vis}-\eta_{inv})^2+(\phi_{vis}-\phi_{inv})^2}$ with $\eta$ being the rapidity.  Putting all $M(\uptau\uptau)$s with the weights into a histogram, the peak value of the histogram is taken as the $M(\uptau\uptau)$ for this event. The probability distributions $\mathcal{P}(\Delta R_{1/2})$ depends upon the momentum of the initial $\uptau$ lepton and the $\uptau$ decay type. They are determined by MC simulations. To consider the effect of the resolution of missing transverse energy, the scanned parameter space is extended to include $\slashed{E}_x$ and $\slashed{E}_y$. The weight will be multiplied by the probability functions $\mathcal{P}(\slashed{E}_x)\times\mathcal{P}(\slashed{E}_y)$. Here $\mathcal{P}(\slashed{E}_{x/y}) = \exp(-(\slashed{E}_{x/y}-\slashed{E}_{x/y}^{obs})^2/2\sigma_{\slashed{E}}^2)$ with $\slashed{E}_{x/y}^{obs}$ being the $x/y$ component of the observed missing transverse energy and $\sigma_{\slashed{E}}$ being the resolution. The method has an efficiency up to $97\%-99\%$ and gives a mass resolution better than that of the CAT.  
\begin{eqnarray}
&& \slashed{E}_x = P_{inv_{1x}} + P_{inv_{2x}} \nonumber \\
&& \slashed{E}_y = P_{inv_{1y}} + P_{inv_{2y}} \nonumber \\
&& m_\uptau = \sqrt{(P_{vis_1}+P_{inv_1})^2 } \nonumber \\
&& m_\uptau = \sqrt{(P_{vis_2}+P_{inv_2})^2 } \label{eq:mmc}
\end{eqnarray}
\item The SVFIT algorithm: The best estimate of $M(\uptau\uptau)$ is taken to be the one maximizing the following probability
\begin{equation}\label{eq:svfit0}
P(M') = \int \delta(M'-M(\vec{X}_{1}, \vec{X}_{2}, \slashed{E}_{x}, \slashed{E}_{y}, P_{vis_{1}},P_{vis_{2}}))\mathcal{P}(\vec{X}_{1}, \vec{X}_{2}, \slashed{E}_{x}, \slashed{E}_{y}, P_{vis_{1}},P_{vis_{2}})d\vec{X}_1d\vec{X}_2
\end{equation}
as a function of the mass hypothesis $M'$. $\vec{X}_{1}$ and $\vec{X}_2$ denote the free paramters to specify the $\uptau$ decay kinematics. $\vec{X}$ is chosen to be $(x_{vis},\phi_{\uptau}, m_{inv})$. Here $x_{vis}$ is the fraction of the $\uptau$ lepton energy carried by the visible decay products in the laboratory frame; $\phi_{\uptau}$ is the azimuthal angle of the $\uptau$ lepton direction in the laboratory frame; $m_{inv}$ is the invariant mass of the neutrino(s) in the $\uptau$ decays. The likelihood $\mathcal{P}(\vec{X}_{1}, \vec{X}_{2}, \slashed{E}_{x}, \slashed{E}_{y}, P_{vis_{1}},P_{vis_{2}})$ is the product of three likelihood functions: two functions model the probability distributions of the decay parameters $X_i$ of the two $\uptau$ leptons and one function quantifies the compatibility of a $\uptau$-pair decay hypothesis with the reconstructed missing transverse energy $\slashed{E}_{x/y}$. The relative $M(\uptau\uptau)$ resolution is (10-20)\%.   
\end{enumerate}

Other interesting methods could be found in Ref.~\cite{3prong, SMR, abhaya, svfitme}. In this paper, we propose an alternative method, which utilizes the momentum component of the $\uptau$ decay products perpendicular to the velocity of the $\uptau$ lepton, $p_\perp$. We will compare this method with the CAT, MMC and SVFIT, as the TMM only reconstructs $M(\uptau\uptau)$ partially. In Sec.~2, we present the principle of this method. In Sec.~3, the method is tested on the MC simulations. In Sec.~4, we apply the selection criteria used by the ATLAS Collaboration in searching for the SM Higgs decaying to the $\uptau$ pair~\cite{htt_atlas}  to a MC sample of $\rm pp\to h(125)+X\to\uptau\uptau+X$ via the gluon-gluon fusion process. The performance of this method would be more realistic in this case. We give the conclusions in Sec.~5. 

\section{\label{sec:principle}Principle of the method}
To illustrate our method of reconstructing the mass of a resonance $R$ decaying to $\uptau^+\uptau-$, we write down the following equations.
	   \begin{eqnarray}
&& M_R = \sqrt{2m_\uptau^2+2E_{\uptau_1}E_{\uptau_2}-2p_{\uptau_1}p_{\uptau_2}\cos\theta_{\uptau_1\uptau_2}} \label{eq:mr} \\
	  && p_{\uptau_{i}} = \sqrt{p_{vis_{i}}^2 + p_{inv_{i}}^2 + 2p_{vis_{i}}p_{inv_{i}}\cos\theta_{vis_{i},inv_{i}}} \label{eq:ptaul} \:, \quad i = 1,2 
	   \end{eqnarray}
Here $M_R$ is the mass of the resonance $R$; $p_{\uptau_{i}}$ is the magnitude of the momentum of the $\uptau$ leptons; $E_{\uptau_{i}}\equiv\sqrt{p_{\uptau_{i}}^2 + m_{\uptau}^2}$; $p_{vis_{i}}$ and $p_{inv_{i}}$ are the magnitude of the momentum of the $\uptau$ decay products, the visible parts (charged leptons or hadronic $\uptau$ jet) and the neutrinos respectively; $\theta_{\uptau_1\uptau_2}$ is the angle between two $\uptau$ leptons; and $\theta_{vis_{i},inv_{i}}$ is the angle between the charged lepton/hadronic jet and the corresponding neutrino(s). To determine the mass of the resonance $M_R$, there are five unknown quantities, $p_{inv_1}$, $p_{inv_2}$, $\theta_{vis_1, inv_1}$, $\theta_{vis_2, inv_2}$ and $\theta_{\uptau_1\uptau_2}$, as shown in Eq.~\ref{eq:mr}-\ref{eq:ptaul}. The other quantities $p_{vis_1}$ and $p_{vis_2}$ can be measured by the detectors.

First of all, we show that $\theta_{\uptau_1\uptau_2}$ can be well estimated by the angle between the visible decay products of two $\uptau$ leptons, $\theta_{vis_1,vis_2}$, in the case of the heavy resonance $R$. We write down the $\uptau$ mass constraints. 
  \begin{eqnarray}
 && m_\uptau^2 = m_{vis}^2+m_{inv}^2+2E_{vis}E_{inv} - 2p_{vis}p_{inv}\cos\theta_{vis,inv} \label{eq:mtau}
 \end{eqnarray}
Here the subscripts $1,2$ are omitted. $m_{vis}$ is the mass of the charged lepton in the $\uptau_l$ mode or the mass of the hadronic jet in the $\uptau_h$ mode; $m_{inv}$ is the mass of the neutrinos in the $\uptau_l$ mode or equal to 0 in the $\uptau_h$ mode; and $E_{vis/inv} \equiv \sqrt{p_{vis/inv}^2 + m_{vis/inv}^2}$. If the resonance is heavy enough $M_R >> m_\uptau$, we have
\begin{equation}
p_{vis}>>m_{vis}, \quad p_{inv} >> m_{inv} \: , \label{eq:colapprox}
\end{equation}
and 
\begin{equation}
E_{vis} \simeq p_{vis}\left( 1+ \frac{m_{vis}^2}{2p_{vis}^2}\right), \quad E_{inv} \simeq p_{inv}\left( 1+ \frac{m_{inv}^2}{2p_{inv}^2}\right) \: . 
\end{equation}
From the $\uptau$ mass constraints shown in Eq.~\ref{eq:mtau}, the angle between visible decay product(s) and the invisible neutrino(s) turns out to be
\begin{eqnarray}
&& \cos\theta_{vis,inv} \simeq 1 + \frac{m_{vis}^2}{2p_{vis}^2}+ \frac{m_{inv}^2}{2p_{inv}^2}-\frac{m_\uptau^2-m_{vis}^2-m_{inv}^2}{2p_{vis}p_{inv}}\label{eq:angle} 
\end{eqnarray}
Using the relation $\cos\theta \simeq 1-\frac{\theta^2}{2}$ for small $\theta$, we find that $|\theta_{vis,inv}|$ is of the order of $\frac{m_{vis}}{p_{vis}}$ or $\frac{m_{inv}}{p_{inv}}$, denoted by $\mathcal{O}(\frac{m_{vis}}{p_{vis}}, \frac{m_{inv}}{p_{inv}})$.  Therefore, in the case of $M_R >> m_\uptau$, the difference between $\theta_{\uptau_1\uptau_2}$ and $\theta_{vis_1,vis_2}$ is of the same order, as shown in Eq.~\ref{eq:xia}.
\begin{equation}\label{eq:xia}
|\theta_{\uptau_1\uptau_2}-\theta_{vis_1,vis_2}| < |\theta_{vis_1,inv_1}|+|\theta_{vis_2,inv_2}| \sim \mathcal{O}(\frac{m_{vis}}{p_{vis}}, \frac{m_{inv}}{p_{inv}})\:.
\end{equation}
Resorting to Eq.~\ref{eq:mr} and Eq.~\ref{eq:xia}, we can estimate the uncertainty of the reconstructed $M_R$ due to the approximation $\theta_{\uptau_1,\uptau_2}\simeq \theta_{vis_1,vis_2}$. 
\begin{equation}\label{eq:xia1}
\frac{\Delta M_R}{M_R} =\left| \frac{\partial M_R}{\partial \theta_{\uptau_1\uptau_2}}\right|\frac{\Delta \theta_{\uptau_1\uptau_2}}{M_R} \sim \frac{p_{\uptau_1}p_{\uptau_2}}{M_R^2}|\sin\theta_{vis_1,vis_2}| \mathcal{O}(\frac{m_{vis}}{p_{vis}}, \frac{m_{inv}}{p_{inv}})  \: ,
\end{equation}
where we have used $\Delta\theta_{\uptau_1\uptau_2}\simeq|\theta_{\uptau_1\uptau_2}-\theta_{vis_1,vis_2}|$. It could be seen that this uncertainty is reduced in the region where $\theta_{vis_1,vis_2}$ is close to $\pi$ (thus $|\sin\theta_{vis_1,vis_2}|$ is close to 0).

If the approximation $\theta_{\uptau_1\uptau_2} \simeq \theta_{vis_1,vis_2}$ is assumed, we then have four unknown quantities in total, namely, $(p_{inv}, m_{inv})$ in the $\uptau_l$ mode, or $(p_{inv}, m_{vis})$ in the $\uptau_h$ mode. The angles $\theta_{vis_1,inv_1}$ and $\theta_{vis_2,inv_2}$ can be expressed as functions of the unknowns through the $\uptau$ mass constraints, Eq.~\ref{eq:mtau}. $M(\uptau\uptau)$ can be determined if $p_{inv}$ and $m_{vis/inv}$ are given. However, $p_{inv}$ is related with the momentum of the $\uptau$ lepton. Thus the probability distribution of $p_{inv}$ depends upon the mass of the mother resonance and is not available for an unknown resonance. Instead, we choose to use the momentum component of the $\uptau$ decay products perpendicular to the velocity of the $\uptau$ lepton, $p_{\perp}$ (it is same for the visible and invisible decay products). $p_{\perp}$ is invariant under the boost in the direction of the $\uptau$ lepton. The joint probability distributions (JPD), $\mathcal{P}(p_{\perp}, m_{vis/inv})$, are then independent upon the mass of the mother resonance.

 To derive the expression of $p_{\uptau}$ as a function of $p_{\perp}$ and $m_{vis/inv}$, let us consider the Lorentz transformation relating the laboratory frame and the center-of-mass (c.m.) frame of the $\uptau$ lepton. The quantities in the  in the c.m. frame will carry a ``$\prime$'' and can be easily calculated for given $m_{inv}$, $m_{vis}$ and $p_{\perp}$. Letting $v$ be the velocity of the $\uptau$ lepton, we define $\beta \equiv v/c$ and $\gamma =1/\sqrt{1-\beta^2}$. We have 
\begin{eqnarray}
&& p_\perp = p_{\perp}^\prime \:, \nonumber \\
&& p_{//vis} \equiv \sqrt{p_{vis}^2 -p_{\perp}^2} = \gamma (p_{//vis}^\prime + \beta E_{vis}^\prime) \:,\nonumber \\
&& E_{vis} \equiv \sqrt{p_{vis}^2 + m_{vis}^2} = \gamma(E_{vis}^\prime+ \beta p_{//vis}^\prime) \:.\label{eq:lorentz}
\end{eqnarray}
In the equations above, $p_{//vis}^\prime = \pm \sqrt{p^{\prime2}-p_{\perp}^2}$ is the momentum component of the visible $\uptau$ decay products parallel to the velocity of the $\uptau$ lepton and $E_{vis}^{\prime2} = \sqrt{p^{\prime2}+m_{vis}^2}$ is the energy of the visible $\uptau$ decay products, where $p^\prime$ is the momentum of the $\uptau$ decay products in the c.m. frame. Using Eqs.~\ref{eq:lorentz} and $p_\uptau = \gamma\beta m_\uptau$, we can obtain 
\begin{equation}
p_{\uptau} = \frac{E_{vis}^\prime p_{//{vis}}\pm E_{vis}p_{//}^\prime}{E_{vis}^{\prime2}-p_{//}^{\prime2}}m_\uptau \:,
\end{equation}
where $p_{//}^\prime \equiv \sqrt{p^{\prime2}-p_{\perp}^2}$  and the two-fold solution is due to $p_{//vis}^\prime = \pm \sqrt{p^{\prime2}-p_{\perp}^2}$ ( physically, it is because the momentum component $p_{//vis}$ in the laboratory frame can be parallel or anti-parallel with that in the c.m. frame of the $\uptau$ lepton).

For convenience, we write down explicitly the equations related with our method.
\begin{eqnarray}
&& M_R \simeq \sqrt{2m_\uptau^2+2E_{\uptau_1}E_{\uptau_2}-2p_{\uptau_1}p_{\uptau_2}\cos\theta_{vis_1,vis_2}} \nonumber \\
&& p_{\uptau_i} = \frac{E_{vis_i}^\prime p_{//{vis_i}}\pm E_{vis_i}p_{//}^\prime}{E_{vis_i}^{\prime2}-p_{//}^{\prime2}}m_\uptau \nonumber \\
&& p_{inv_i} = \sqrt{(E_{\uptau_i}-E_{vis_i})^2 - m_{inv}^2}\label{eq:key}
\end{eqnarray}
where the subscript $i = 1, 2$ and $\theta_{\uptau_1\uptau_2}$ is replaced by $\theta_{vis_1,vis_2}$ in the first equation. 

The next step is to sample all possible values of $(p_{\perp}, m_{inv})$ for the $\uptau_l$ mode or $(p_{\perp}, m_{vis})$ for the $\uptau_h$ mode according to their JPDs which can be determined by the MC simulations. For each event accumulated by the detectors, we sample the unknowns many times (10000 is fairly enough) and obtain a distribution of $M(\uptau\uptau)$ according to Eqs.~\ref{eq:key}. We assume that the mass of the resonance lies at the position with the maximal probability. 
For each sampling entry, the two-fold solutions, shown in the second line of Eqs.~\ref{eq:key}, are used as long as they are physically allowed, namely, $p_{\uptau_i}>0$.

To use the missing transverse energy, each sampling entry is given a weight, $w(\slashed{E})$, which is defined as
\begin{equation}~\label{eq:wmet}
w(\slashed{E}) = \frac{1}{\sqrt{2\pi}\sigma_{\slashed{E}_x}}e^{-\frac{(\slashed{E}_x^{obs}-\slashed{E}_x)^2}{2\sigma_{\slashed{E}_x}^2}}\times \frac{1}{\sqrt{2\pi}\sigma_{\slashed{E}_y}}e^{-\frac{(\slashed{E}_y^{obs}-\slashed{E}_y)^2}{2\sigma_{\slashed{E}_y}^2}} \: .
\end{equation}
Here ${\slashed{E}}_{x/y}^{obs}$ are the $x/y$ components of the observed missing energy; $\sigma_{\slashed{E}_x}$ and $\sigma_{\slashed{E}_y}$ are the corresponding resolutions; and $\slashed{E}_{x/y}$ are the $x/y$ components of the sum of the momenta of the neutrinos, shown in Eq.~\ref{eq:met}.
\begin{eqnarray}
\slashed{E}_x = \sum_{i=1,2}p_{inv_i}\sin\theta_{vis_i}\cos\phi_{vis_i} \nonumber \\
\slashed{E}_y = \sum_{i=1,2}p_{inv_i}\sin\theta_{vis_i}\sin\phi_{vis_i}  \label{eq:met}
\end{eqnarray}
Here the collinear approximation is used; $\theta_{vis_i}$ and $\phi_{vis_i}$ are the polar angle and the azimuthal angle of the visible decay products respectively. To comply to the experimental performance~\cite{met_atlas, met_cms}, the missing energy resolution is parameterized in the form,
\begin{equation}
\sigma_{\slashed{E}_x} = \sigma_{\slashed{E}_y} = 0.7\times \sqrt{\sum E_T}  \:,
\end{equation}
where $\sum E_T$ is the scalar sum of all the observed objects  (electrons, muons, hadronic taus, jets, etc.) and defined as 
\begin{equation}\label{eq:met_def}
\sum E_T \equiv  \sum_{\text{observed objects}} \sqrt{p_x^2+p_y^2}\:.
\end{equation}

\section{Performance of the technique}\label{sec:performance}
To test the performance of this technique, the MC samples for the physics processes $\rm pp\to Z/h(125)/h(750)+X\to \uptau^+\uptau^-+X$ at 13~TeV are produced with MadGraph5~\cite{madgraph}. The parton showers are simulated by Pythia~8~\cite{pythia} and the detector response is simulated by Delphes~3~\cite{delphes}. The detector simulation is adjusted to meet the run 1 performance of the ATLAS detector. Most relevantly, the reconstruction efficiency of the hadronic $\uptau$ jets is about 60\% with a jet faking rate about 1\%. The effect of the pileup interactions is not considered until the end of this section.  Here $\rm h(125)$ denotes the SM Higgs boson with the mass about 125~GeV/$c^2$~\cite{higgs_mass} and the width 4.07~MeV. $\rm h(750)$ denotes a possible high-mass Higgs boson. Its mass and width are 750~GeV/$c^2$ and 40~GeV respectively in the simulations. 

In the leptonic decay mode $\uptau_l$, we do not distinguish the charged leptons $\rm e$ and $\upmu$ as their masses are negligible compared to the $\uptau$ mass. In the hadronic decay mode $\uptau_h$, the events with one charged track and three charged tracks are considered separately. Events with two $\uptau$ candidates, charged leptons or hadronic $\uptau$ jets, with opposite charge sign are selected. The transverse momentum, $p_T$, is required to be larger than 20~GeV/$c$ and  the rapidity $|\eta|$ is required to be less than 2.5 for both $\uptau$ candidates. They are further required to be isolated. The charged leptons satisfy $p_T(\Delta R=0.5)/p_T(l)<10\%$. Here  $p_T(l)$ is the transverse momentum of the lepton and $p_T(\Delta R=0.5)$ is the transverse momentum of the additional observed objects in the cone around the lepton candidate, where the cone size is defined by the angular distance between the additional objects and the lepton candidate $\Delta R=0.5$. For a hadronic $\uptau$ candidate, the angular distance with respect to an electron or a muon is required to be larger than 0.2. Experimentally, additional isolation conditions are imposed on the hadronic $\uptau$ candidates to reduce the jet faking rate. For example, the ATLAS collaboration uses the discriminating variables based on the tracks with $p_T>1$~GeV/$c$ and the energy deposited in the calorimeter in the cone $\Delta R<0.2$ and those in the region $0.2<\Delta R<0.4$ around the hadronic $\uptau$ candidate's direction~\cite{htt_atlas}. Here in the delphes simulation, no further isolation condition is used for the hadronic $\uptau$s. 

\begin{figure}[htbp]
\center
\includegraphics[width = 0.50\textwidth]{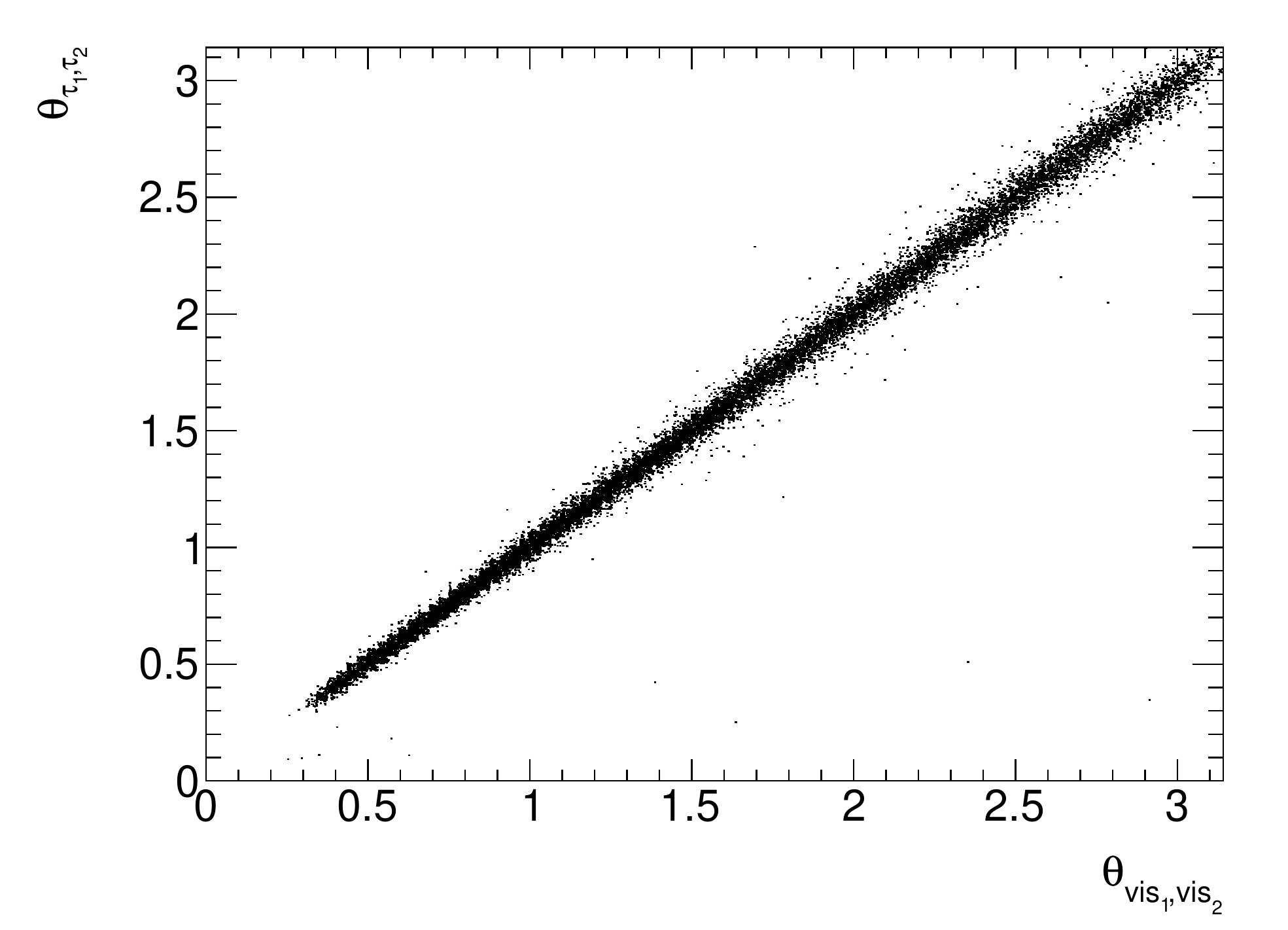}
\put(-190, 150){\textbf{(a)}}
\includegraphics[width = 0.50\textwidth]{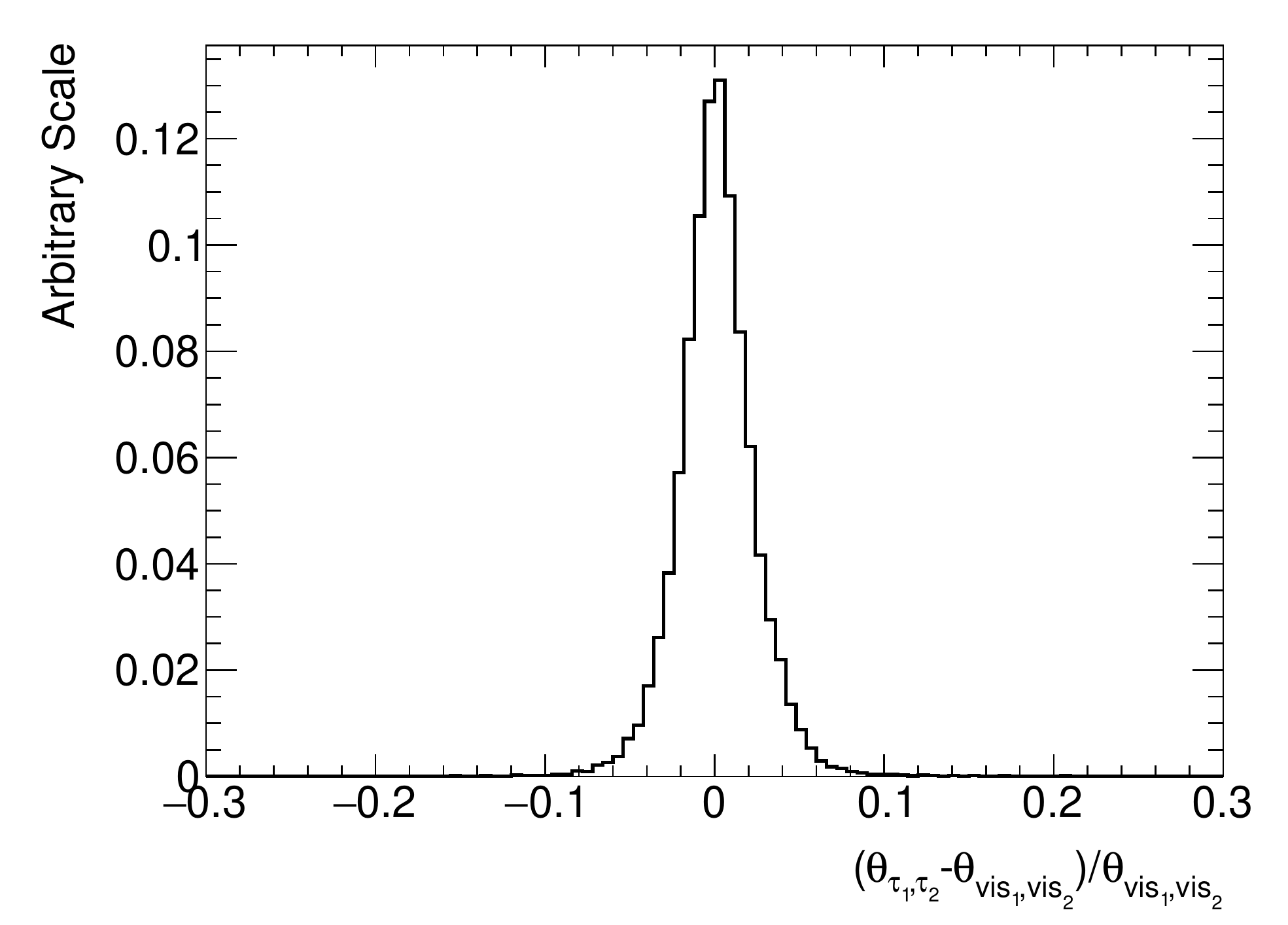}
\put(-190, 150){\textbf{(b)}}
\caption{\label{fig:colapprox} (a) The distribution of the angle between two $\uptau$ leptons, $\theta_{\uptau_1\uptau_2}$, versus the angle between the two visible $\uptau$ candidates, $\theta_{vis_1,vis_2}$. (b) The distribution of $(\theta_{\uptau_1\uptau_2}-\theta_{vis_1,vis_2})/\theta_{vis_1,vis_2}$. The plots are from the MC simulation of $\rm Z\to\uptau\uptau$.} 
\end{figure}

In the first place, the collinear approximation that the angle between the $\uptau$ leptons is well estimated by the angle between the two visible $\uptau$ candidates is verified in the MC simulations. Figure~\ref{fig:colapprox}~(a) shows the distribution of $\theta_{\uptau_1\uptau_2}$ versus $\theta_{vis_1,vis_2}$ while Fig.~\ref{fig:colapprox}~(b) shows the distribution of $(\theta_{\uptau_1,\uptau_2}-\theta_{vis_1,vis_2})/\theta_{vis_1,vis_2}$. We find that the correlation coefficient between the two angles is 0.998 and the relative difference $|\theta_{\uptau_1,\uptau_2}-\theta_{vis_1,vis_2}|/\theta_{vis_1,vis_2}$ is well below 10\%. To be exact, the mean value of the distribution in Fig.~\ref{fig:colapprox}~(b) is 0.1\% and the root mean square (RMS) is 2.6\%.

\begin{figure}[htbp]
\center
\includegraphics[width = 0.50\textwidth]{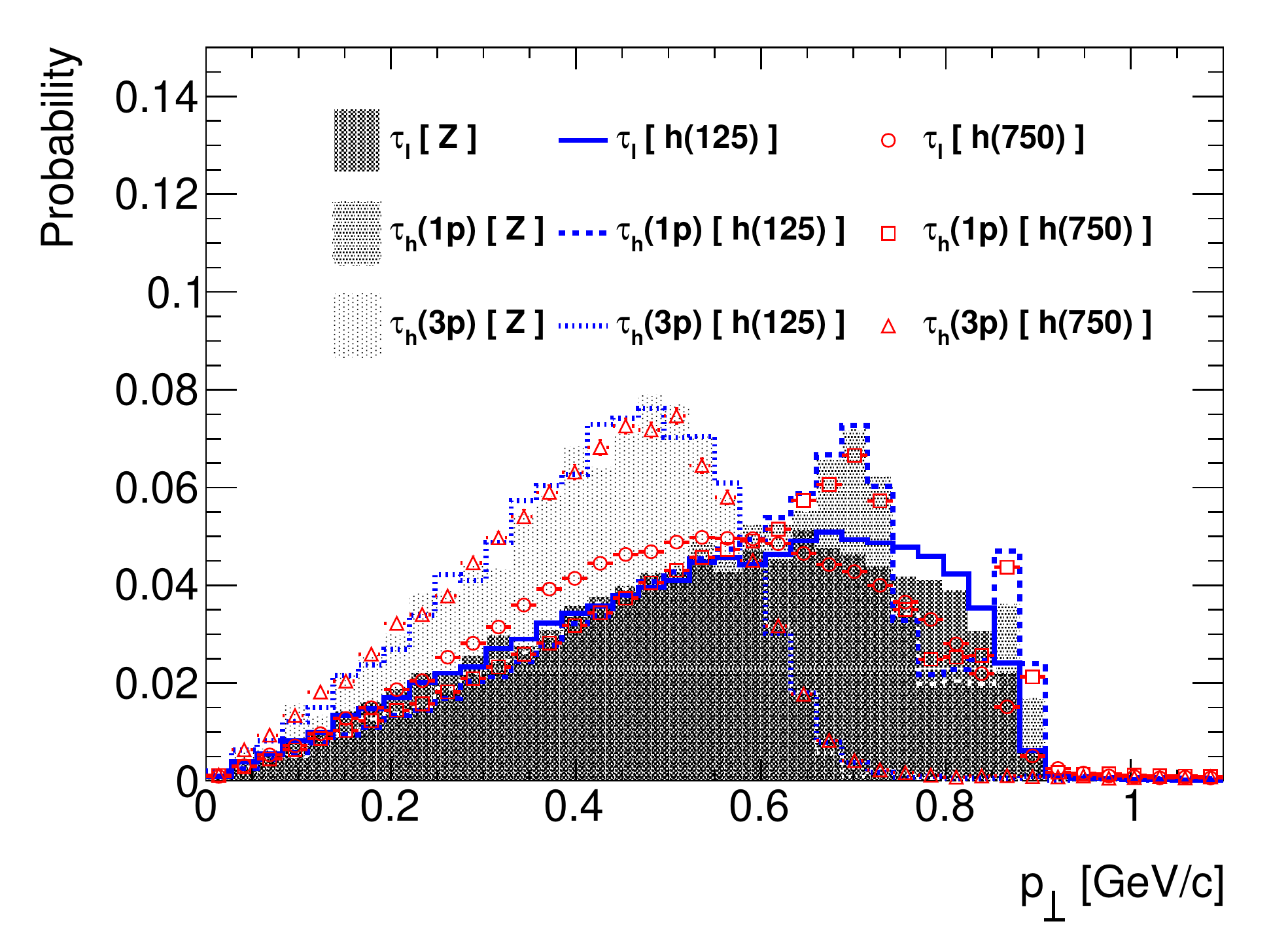}
\put(-210, 150){\textbf{(a)}}
\includegraphics[width = 0.50\textwidth]{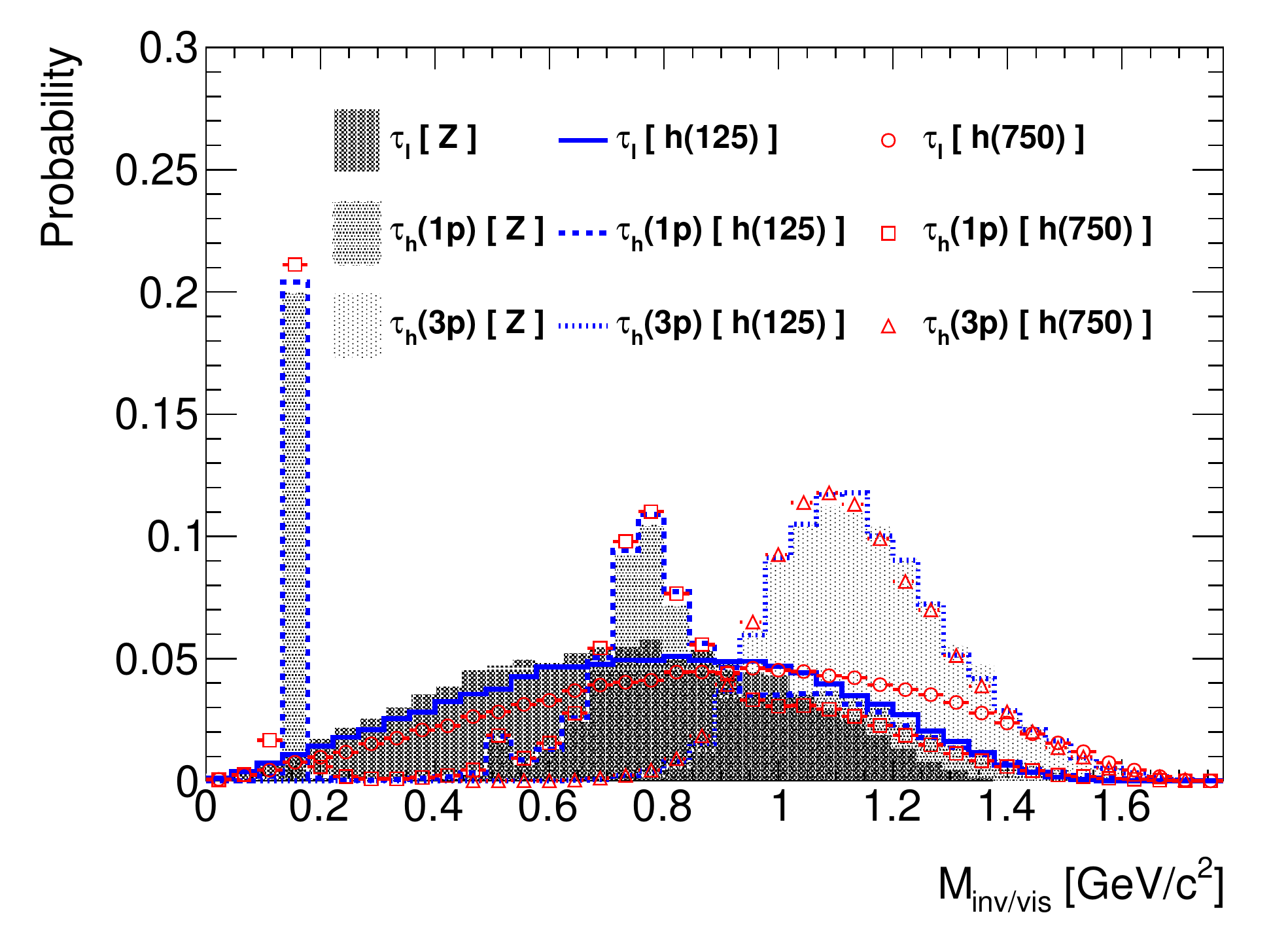}
\put(-210, 150){\textbf{(b)}}
\caption{\label{fig:invp_Mvis} (color online) (a) The distributions of $p_\perp$ in the lepton mode and the hadron mode with one charged track or three charged tracks. (b) The mass distributions of the invisible neutrinos in the lepton mode and the visible hadrons in the hadron mode with one charged track or three charged tracks. The hatched histograms/curves/markers represent the distributions for the resonance $\rm Z/h(125)/h(750)$. All distributions are normalized to a unit area. } 
\end{figure}

\begin{figure}[htbp]
\center
\includegraphics[width = 0.50\textwidth]{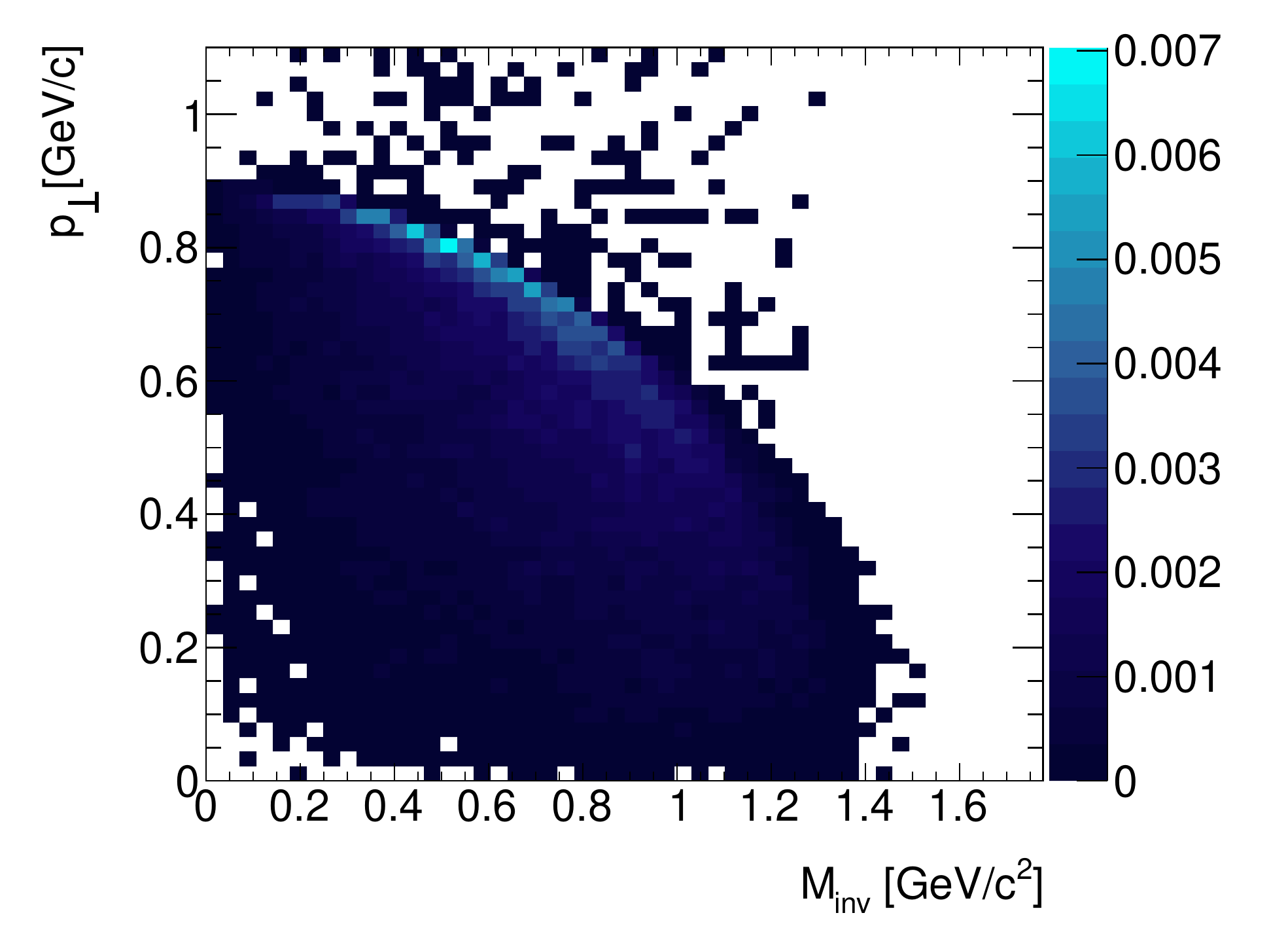}
\put(-70, 150){\textbf{(a)}}
\includegraphics[width = 0.50\textwidth]{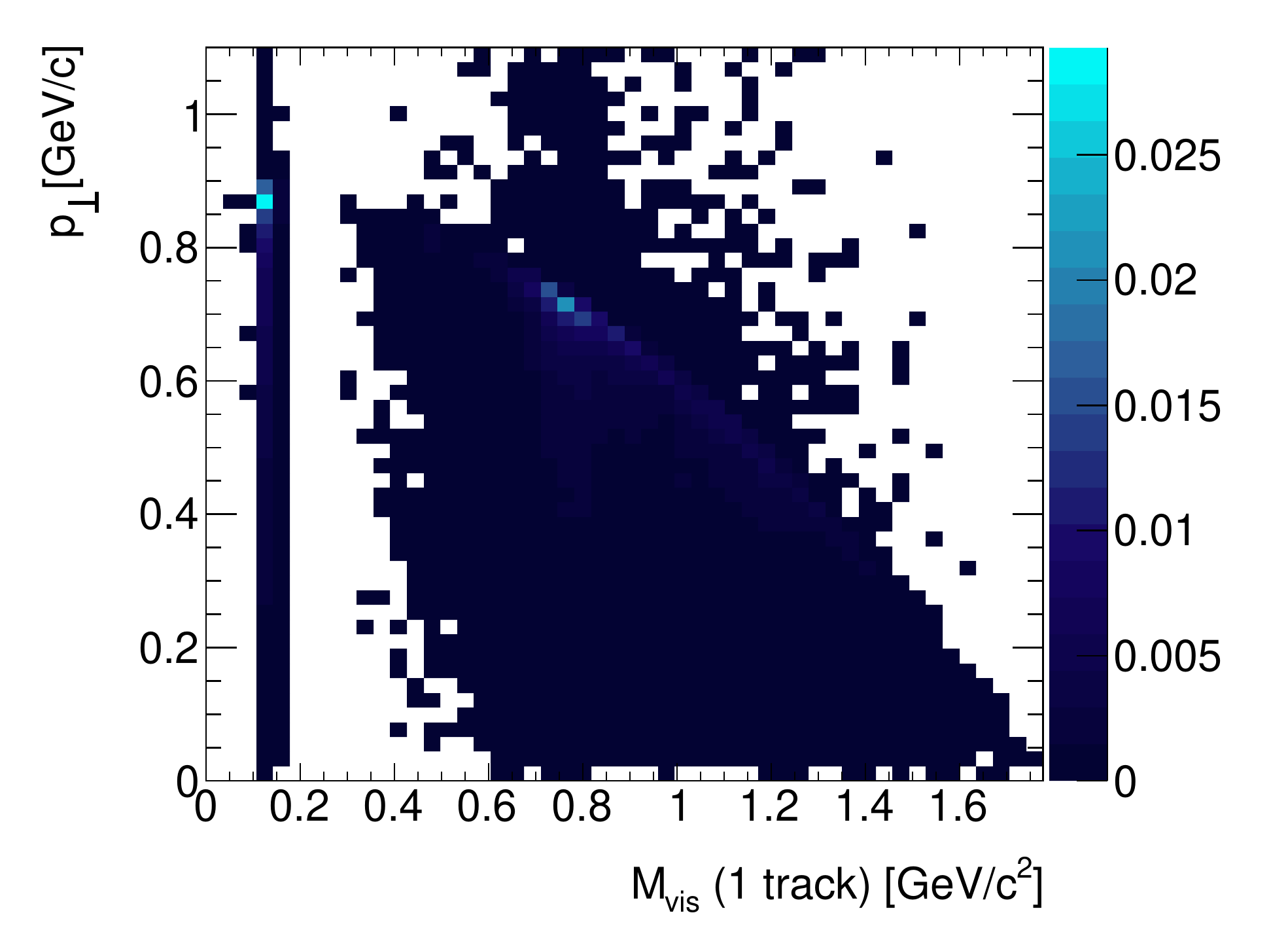}
\put(-70, 150){\textbf{(b)}}\\
\includegraphics[width = 0.50\textwidth]{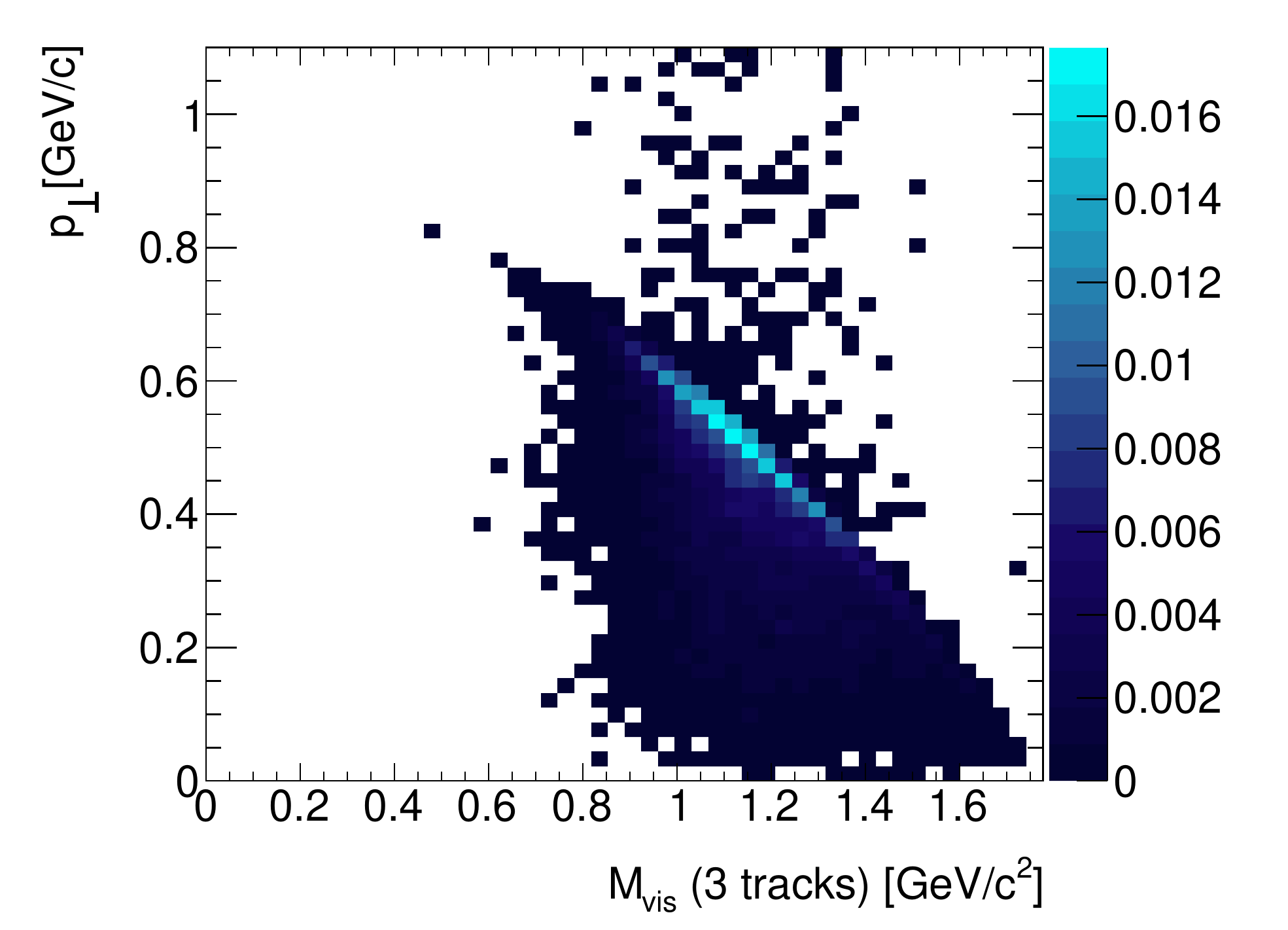}
\put(-70, 150){\textbf{(c)}}
\caption{(color online)\label{fig:invp_Mvis2D} (a) shows the distribution of $p_{\perp}$ versus $m_{inv}$ in the $\uptau_l$ mode; (b) and (c) show the distributions of $p_{\perp}$ versus $m_{vis}$ in the $\uptau_h$ mode with one charged track and three charged tracks respectively. All distributions are normalized to a unit volume. The plots are from the MC simulation of $\rm Z\to\uptau\uptau$. }
\end{figure}

Figure~\ref{fig:invp_Mvis} shows the distributions of the momentum component $p_\perp$, the invisible mass distribution in the $\uptau_l$ mode and the visible mass distribution in the $\uptau_h$ mode for the three mother resonances. In Fig.~\ref{fig:invp_Mvis}~(b), the hadronic modes with one charged track, $\rm \uptau\to\uppi/K/\uprho+\upnu_\uptau$, can be well recognized. We see that the probability distributions have little dependence upon the mass of the mother resonance.
$p_{\perp}$ and $m_{vis/inv}$ are correlated. Figure~\ref{fig:invp_Mvis2D} shows the JPDs $\mathcal{P}(p_\perp, m_{inv})$ in the $\uptau_l$ model and $\mathcal{P}(p_{\perp}, m_{vis})$ in the $\uptau_h$ mode. 

\begin{figure}[htbp]
\center
\includegraphics[width = 0.50\textwidth]{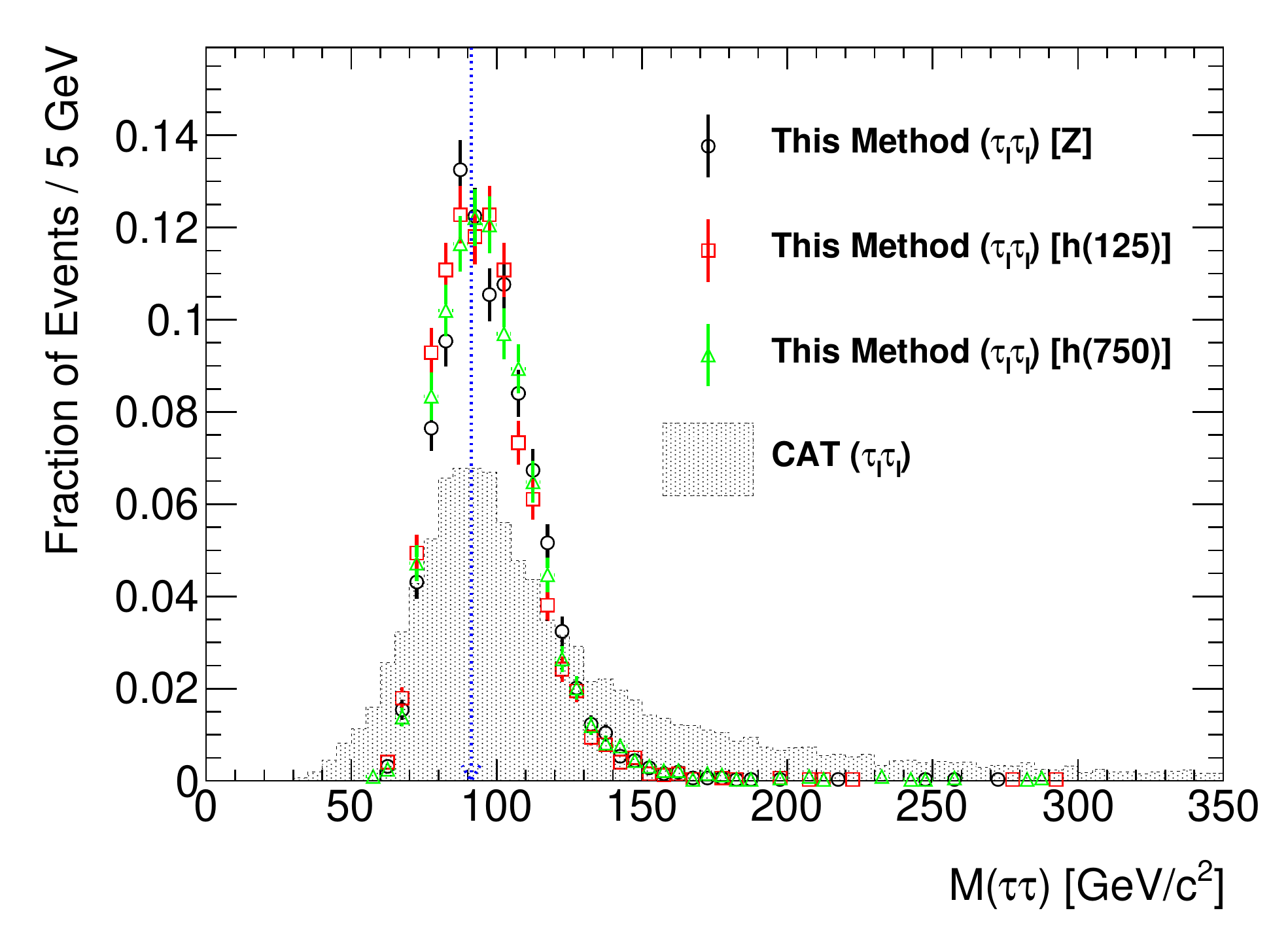}
\put(-190, 150){\textbf{(a)}}
\includegraphics[width = 0.50\textwidth]{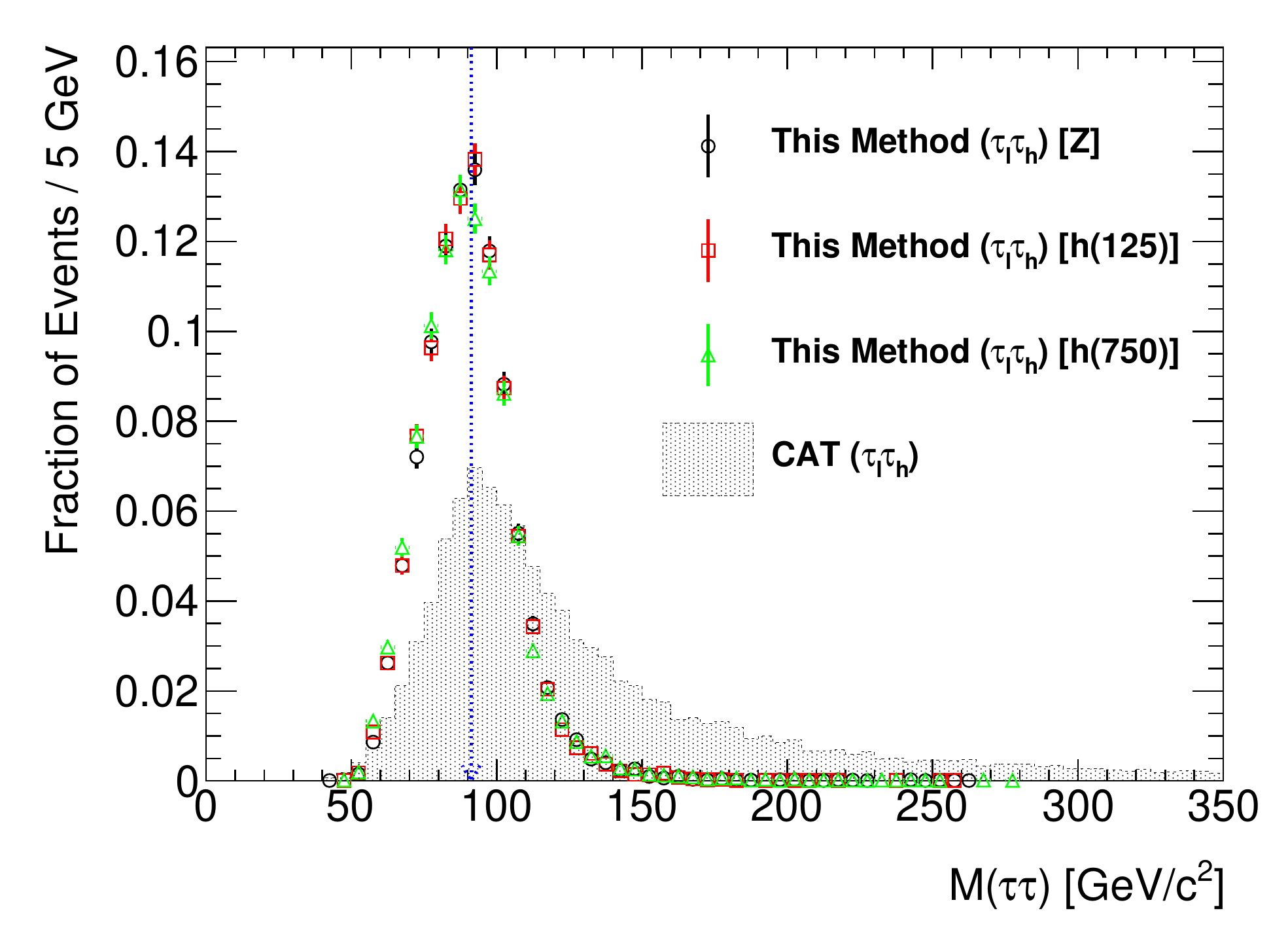}
\put(-190, 150){\textbf{(b)}}\\
\includegraphics[width = 0.50\textwidth]{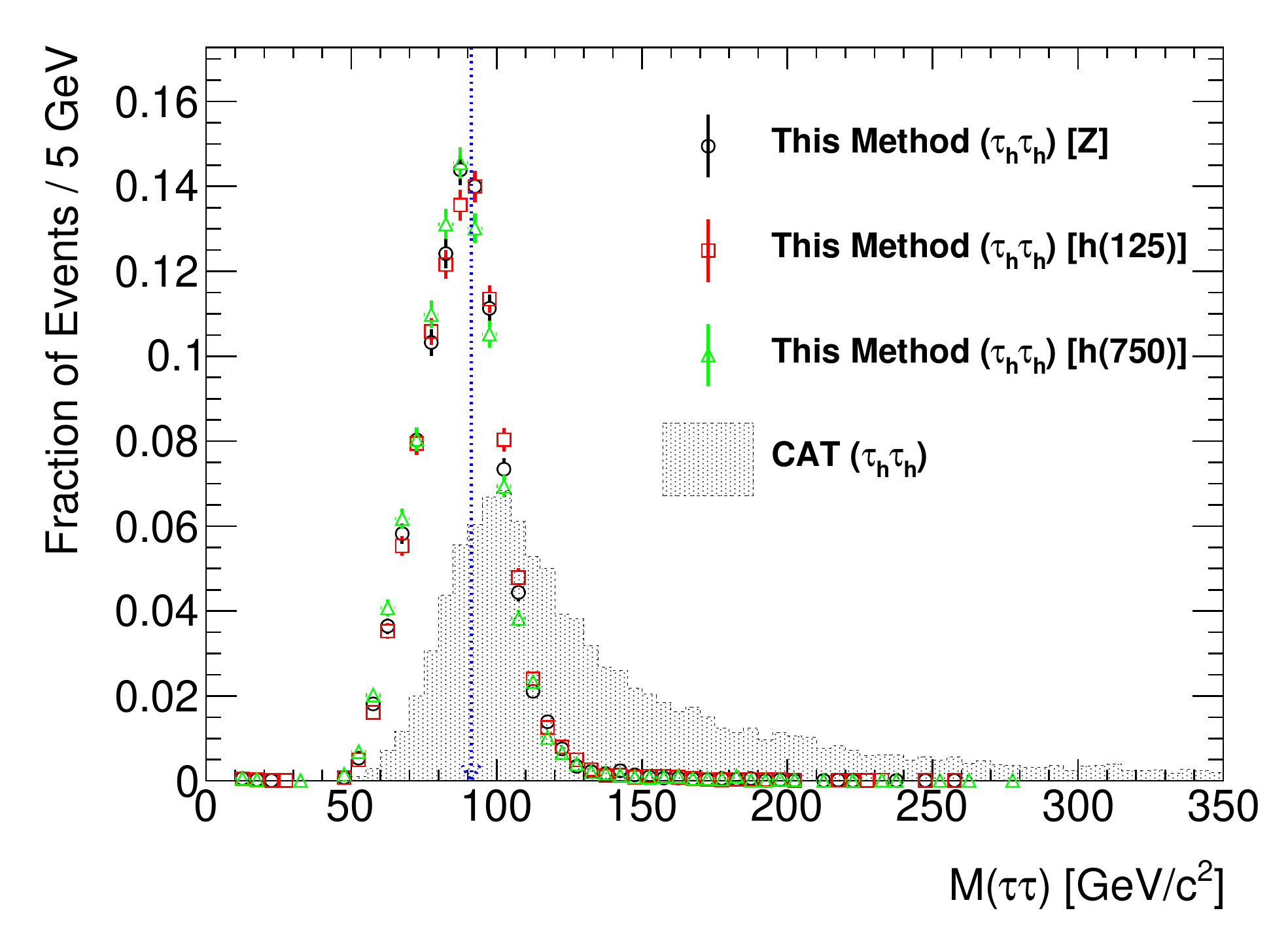}
\put(-190, 150){\textbf{(c)}}
\caption{(color online)\label{fig:zmtt} The distributions of $M(\uptau\uptau)$ for $\rm Z\to\uptau^+\uptau^-$ in the $\uptau_l\uptau_l$ mode (a), the $\uptau_l\uptau_h$ mode (b), and the $\uptau_h\uptau_h$ mode (c). The open circles, boxes and triangles represent the results in this work with the joint probability distributions from the simulations for the resonance $\rm Z$, $\rm h(125)$ and $\rm h(750)$, respectively.  The hatched histograms are the results using the collinear approximation technique (CAT). All distributions are normalized to a unit area. The blue dashed arrow in each plot denotes the position of the generated mass.} 
\end{figure}
\begin{figure}[htbp]
\center
\includegraphics[width = 0.50\textwidth]{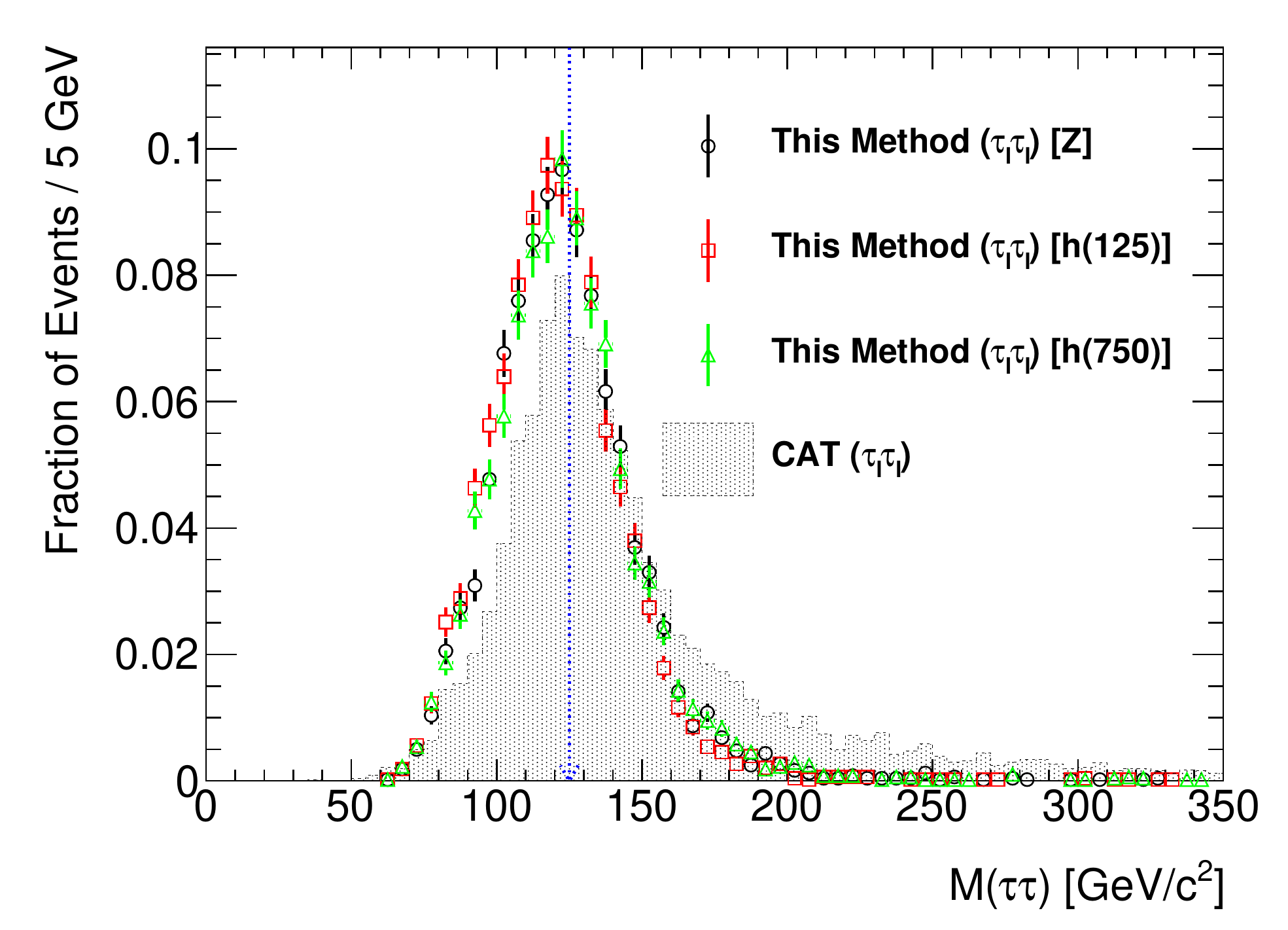}
\put(-190, 150){\textbf{(a)}}
\includegraphics[width = 0.50\textwidth]{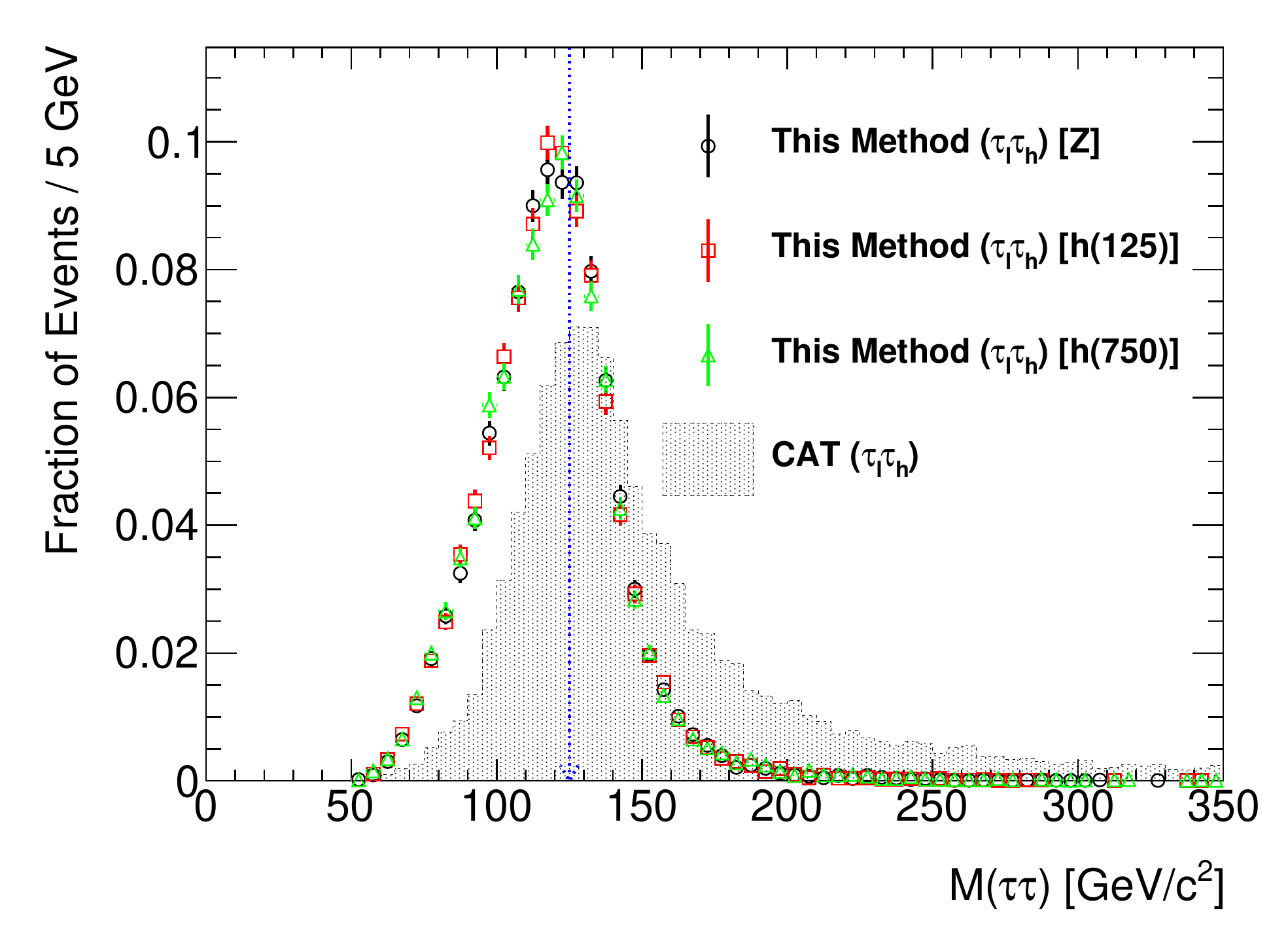}
\put(-190, 150){\textbf{(b)}}\\
\includegraphics[width = 0.50\textwidth]{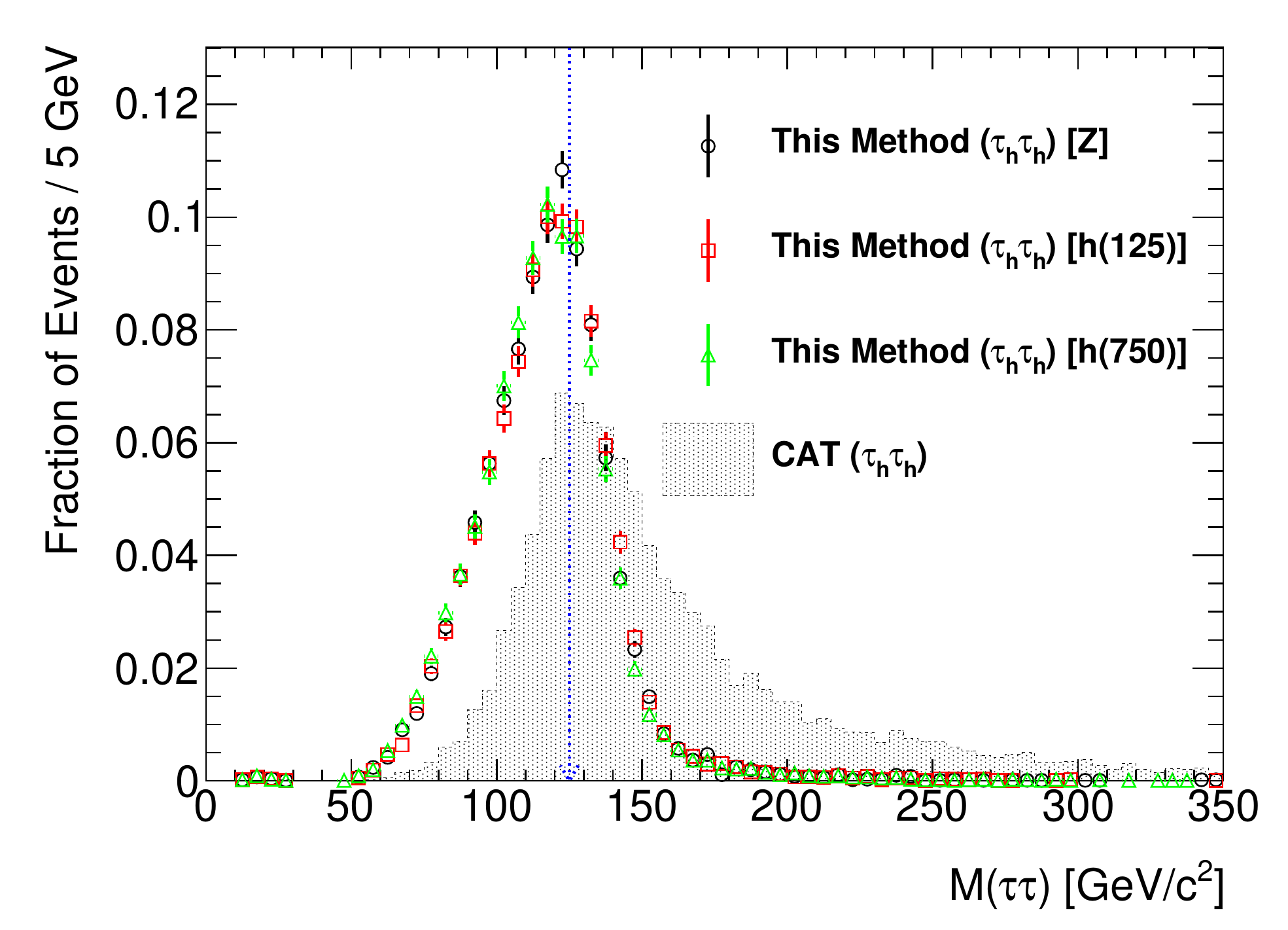}
\put(-190, 150){\textbf{(c)}}
\caption{(color online)\label{fig:hmtt} The distributions of $M(\uptau\uptau)$ for $\rm h(125)\to\uptau^+\uptau^-$ in the $\uptau_l\uptau_l$ mode (a), the $\uptau_l\uptau_h$ mode (b), and the $\uptau_h\uptau_h$ mode (c). The open circles, boxes and triangles represent the results in this work with the joint probability distributions from the simulations for the resonance $\rm Z$, $\rm h(125)$ and $\rm h(750)$, respectively.  The hatched histograms are the results using the collinear approximation technique (CAT). All distributions are normalized to a unit area. The blue dashed arrow in each plot denotes the position of the generated mass.} 
\end{figure}
\begin{figure}[htbp]
\center
\includegraphics[width = 0.50\textwidth]{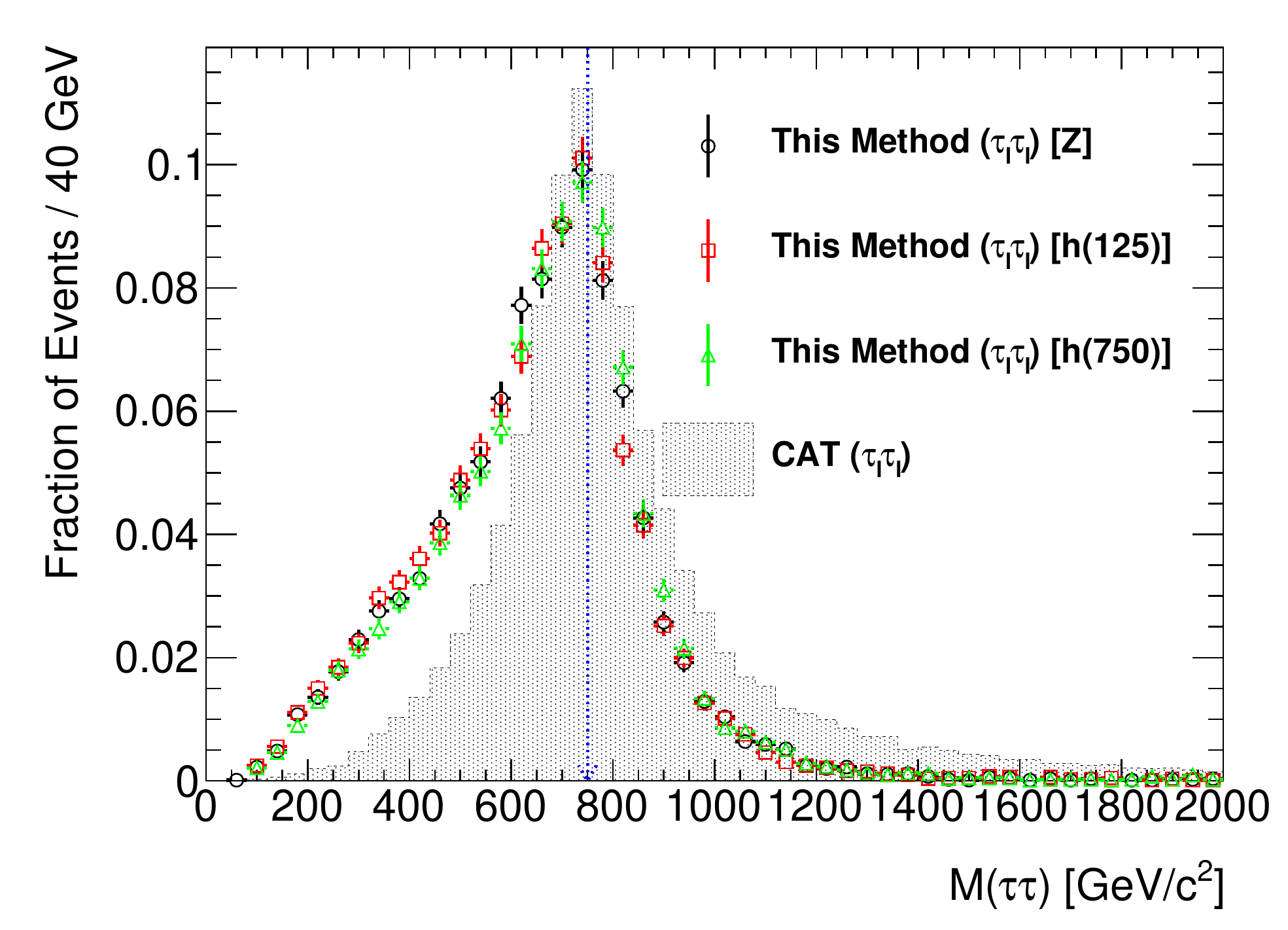}
\put(-190, 150){\textbf{(a)}}
\includegraphics[width = 0.50\textwidth]{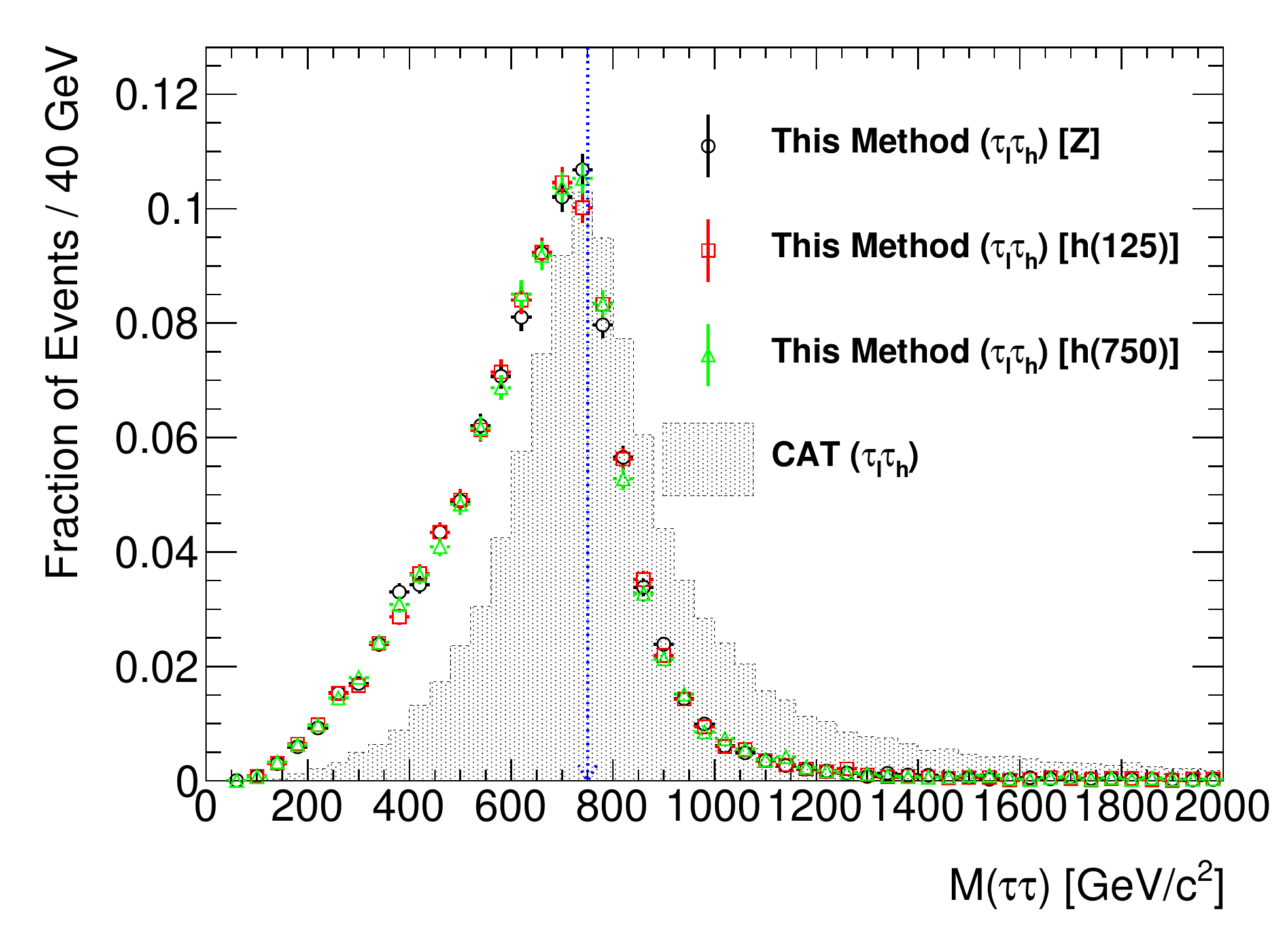}
\put(-190, 150){\textbf{(b)}}\\
\includegraphics[width = 0.50\textwidth]{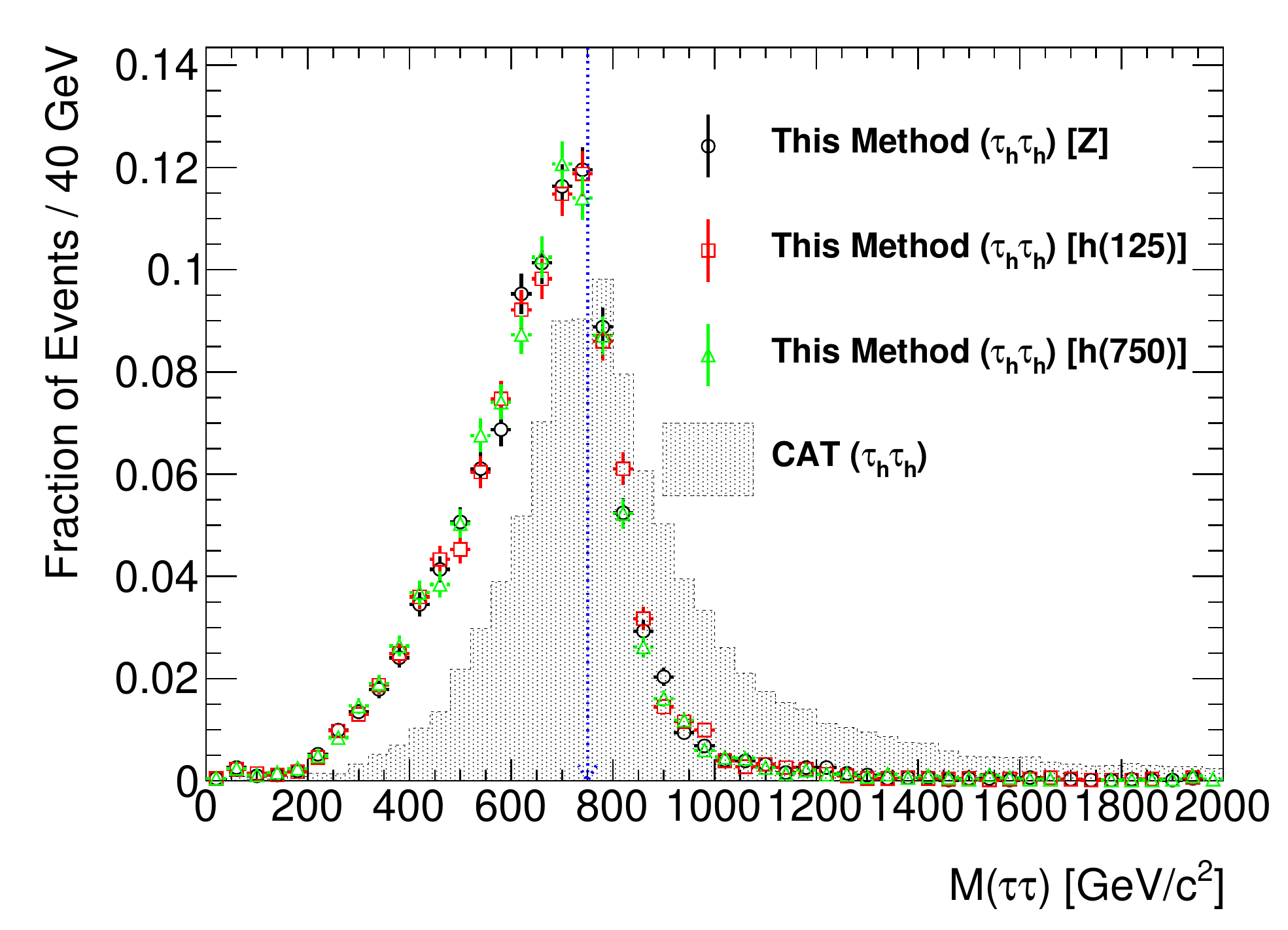}
\put(-190, 150){\textbf{(c)}}
\caption{(color online)\label{fig:hpmtt} The distributions of $M(\uptau\uptau)$ for $\rm h(750)\to\uptau^+\uptau^-$ in the $\uptau_l\uptau_l$ mode (a), the $\uptau_l\uptau_h$ mode (b), and the $\uptau_h\uptau_h$ mode (c). The open circles, boxes and triangles represent the results in this work with the joint probability distributions from the simulations for the resonance $\rm Z$, $\rm h(125)$ and $\rm h(750)$, respectively.  The hatched histograms are the results using the collinear approximation technique (CAT). All distributions are normalized to a unit area. The blue dashed arrow in each plot denotes the position of the generated mass.} 
\end{figure}

For each event, we sample $(p_\perp, m_{vis/inv})$ according to their JPDs for 10000 times. A distribution of $M(\uptau\uptau)$ is obtained using Eqs.~\ref{eq:key}. The best estimate of $M(\uptau\uptau)$ for this event is assumed to be the peak position of this distribution. The final reconstructed distributions of $M(\uptau\uptau)$ for $\rm Z/h(125)/h(750)\to\uptau^+\uptau^-$ are shown in Fig.~\ref{fig:zmtt}, Fig.~\ref{fig:hmtt} and Fig.~\ref{fig:hpmtt} respectively. For each resonance, we also present the results based on the JPDs $\mathcal{P}(p_\perp, m_{vis/inv})$ from the simulation of the other resonances. In fact, the JPDs describe the decay kinematics of the $\uptau$ lepton and have nothing to do with the resonance decaying to $\uptau$ lepton pairs. Therefore, we can reconstruct the mass of an unknown resonances based on the JPDs from an known resonance. As Fig.~\ref{fig:zmtt}, Fig.~\ref{fig:hmtt} and Fig.~\ref{fig:hpmtt} shows, the performance are nearly the same no matter what JPDs are used.

We adopt two quantities to measure the performance, namely, the reconstruction efficiency due to the technique itself and the relative mass resolution.  They are crucial elements to search for new resonances and distinguish the resonance signal from the background. They are explained below. 

The reconstruction efficiency describes the rate of successful mass reconstruction and is defined  as 
\begin{equation}\label{eq:eff}
\epsilon_{success} = \frac{N_{sel+success}}{N_{sel}} \: ,
\end{equation}
where $N_{sel}$ is the number of events passing the selection criteria, and $N_{sel+success}$ is the number of events which are successfully reconstructed from the selected events.
In the CAT, the mass can not be reasonably reconstructed if $x_{1/2}<0$ or $x_{1/2}>1$ in Eq.~\ref{eq:cat}. The efficiency is about 40\%-70\%. In the MMC, the efficiency loss is  only 1\%. It is due to large fluctuations of the $\slashed{E}_T$ measurement or other scan variables and the limited number of scans.

The definition of the relative mass resolution, used in Ref.~\cite{htt_atlas}, is the ratio of the full width at half maximum (FWHM) and the peak value ($m_{peak}$) of the mass distribution, denoted by FWHM/$m_{peak}$.    

The comparison of the performances of the CAT and this method is summarized in Table~\ref{tab:compare}.  If the resonance is heavier, the collinear approximation is better and thus the CAT works better. Our method has a stable performance with no efficiency lose and give a relative mass resolution of (30-40)\%. 

For the present, we have not yet mentioned the backgrounds under this method. In searching for the SM higgs decay $\rm h(125) \to\uptau^+\uptau^-$ for which only evidences are reported~\cite{svfit,htt_atlas}, the decay $\rm Z\to\uptau^+\uptau^-$ is the dominant background since the masses of the two bosons are close. A $M(\uptau\uptau)$ reconstruction technique with a higher reconstruction efficiency and a better mass resolution will surely improve the signal-background separation and increase the signal significance. Another important background is the multi-jet background which dominates the faking hadronic $\uptau$s. It is usually estimated by a data-driven method, namely, it is represented by the data events with two $\uptau$ candidates having the same charge sign. 

\begin{figure}[htbp]
\center
\includegraphics[width = 0.50\textwidth]{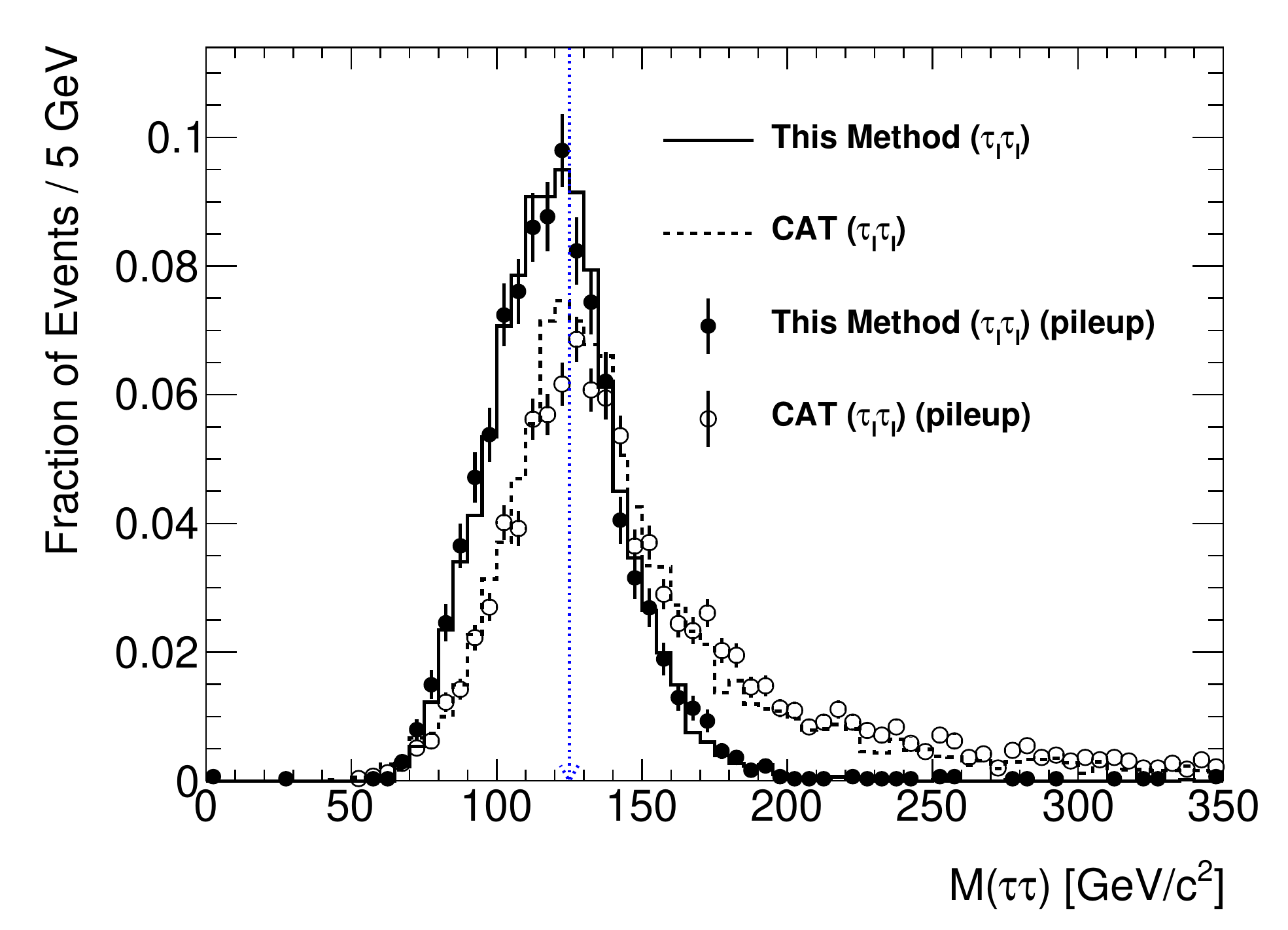}
\put(-190, 150){\textbf{(a)}}
\includegraphics[width = 0.50\textwidth]{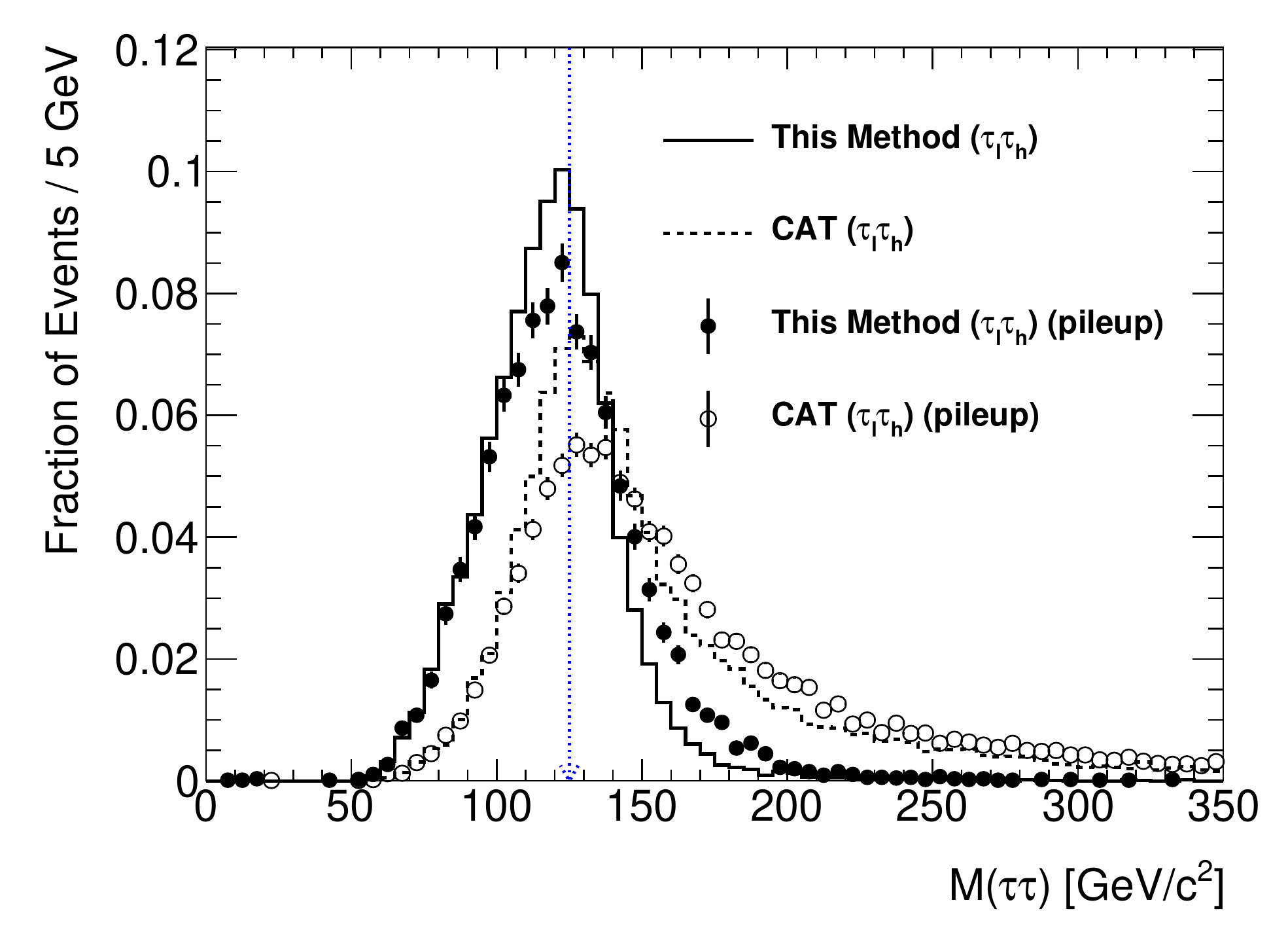}
\put(-190, 150){\textbf{(b)}}\\
\includegraphics[width = 0.50\textwidth]{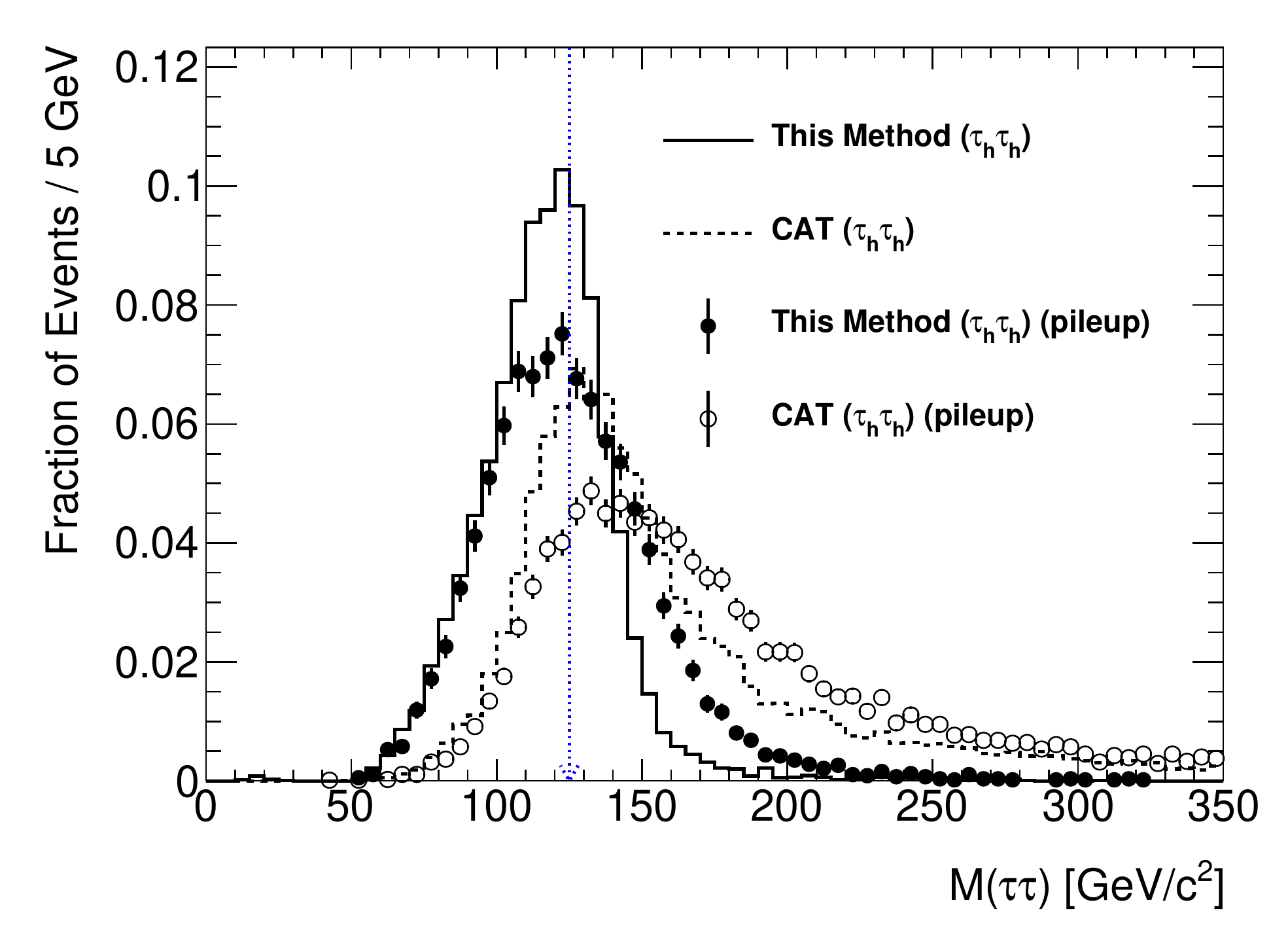}
\put(-190, 150){\textbf{(c)}}
\caption{(color online)\label{fig:pileup} The distributions of $M(\uptau\uptau)$ for $\rm h(125)\to\uptau^+\uptau^-$ in the $\uptau_l\uptau_l$ mode (a), the $\uptau_l\uptau_h$ mode (b), and the $\uptau_h\uptau_h$ mode (c). The solid histograms (solid dots with error bar) represent the results in this work without (with) considering the pileup interactions, while the dashed histograms and the open circles with error bar represent the results using the collinear approximation technique (CAT). All distributions are normalized to a unit area. The blue dashed arrow in each plot denotes the position of the generated mass.} 
\end{figure}

In the end of this section, we take $\rm h(125) \to\uptau\uptau$ as example to investigate the effect of the pileup interactions. Here the average pileup is assumed to be 50. By repeating the simulations with the pileup interactions considered, it is found the mass reconstruction performance become worse, especially for the $\uptau_h\uptau_h$, as shown in Fig.~\ref{fig:pileup}. In this case, however, our method still give better mass resolutions than the CAT. The results are summarized in the last two lines of Table~\ref{tab:compare}.
It should be noted that very simple selection requirements are used in this section. In next section, we will test this method in a more realistic case.

\begin{table}[htbp]
\center{}
\caption{(color online)\label{tab:compare} Performance comparison of the mass reconstruction method in this work and the collinear approximation technique (CAT). The last two lines summarize the results considering the effect of the pileup interactions.}
\begin{tabular}{l l l l l l l l }
\hline
\multirow{2}{*}{Technique} &\multirow{2}{*}{Resonance} & \multicolumn{3}{c}{Efficiency} &   \multicolumn{3}{c}{FWHM/$m_{peak}$} \\ 
& & $\uptau_l\uptau_l$ & $\uptau_l\uptau_h$ & $\uptau_h\uptau_h$ & $\uptau_l\uptau_l$ & $\uptau_l\uptau_h$ & $\uptau_h\uptau_h$\\
\hline
CAT & $\rm Z$ & 51\%&48\%&45\% & 46\%&46\%&49\%\\
& $\rm h(125)$ & 61\%&55\%&51\% & 33\%&35\%&39\% \\
& $\rm h(750)$ & 66\%&59\%&52\% & 32\%&32\%&38\% \\
\hline
This work & $\rm Z$ & \multicolumn{3}{c}{100\%} & 34\%&34\%&32\%\\
& $\rm h(125)$ &  \multicolumn{3}{c}{100\%} & 33\%&33\%&33\%\\
& $\rm h(750)$ & \multicolumn{3}{c}{100\%} & 38\%&38\%&32\% \\
\hline
CAT (pileup) & $\rm h(125)$ &51\%&48\%&45\%& 39\%&55\%&60\%\\
This work (pileup) & $\rm h(125)$ & \multicolumn{3}{c}{100\%} & 33\%&37\%&49\%\\
\hline
\end{tabular}
\end{table}

\section{Performance using the ATLAS selection criteria}\label{sec:atlas}
In this section, we apply the selection criteria used by the ATLAS Collaboration in searching for the SM Higgs boson decaying to the $\uptau$ pair~\cite{htt_atlas} to a MC sample of $\rm pp\to h(125)+X\to\uptau\uptau+X$ via the gluon-gluon fusion process at 8~TeV. In this case, the performance of this technique would be more realistic.

\begin{figure}[htbp]
\center
\includegraphics[width = 0.50\textwidth]{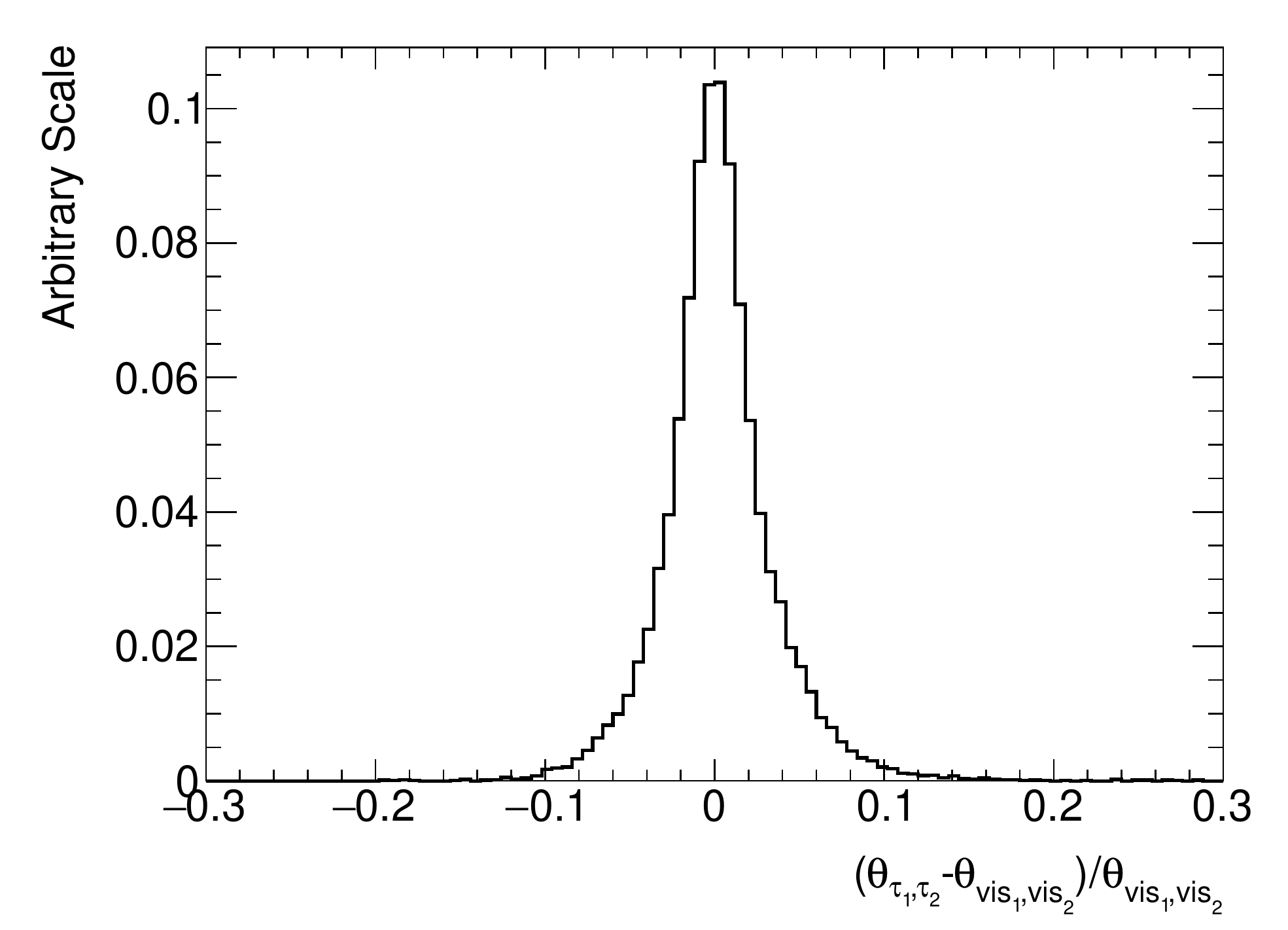}
\put(-190, 150){\textbf{(a)}}
\includegraphics[width = 0.50\textwidth]{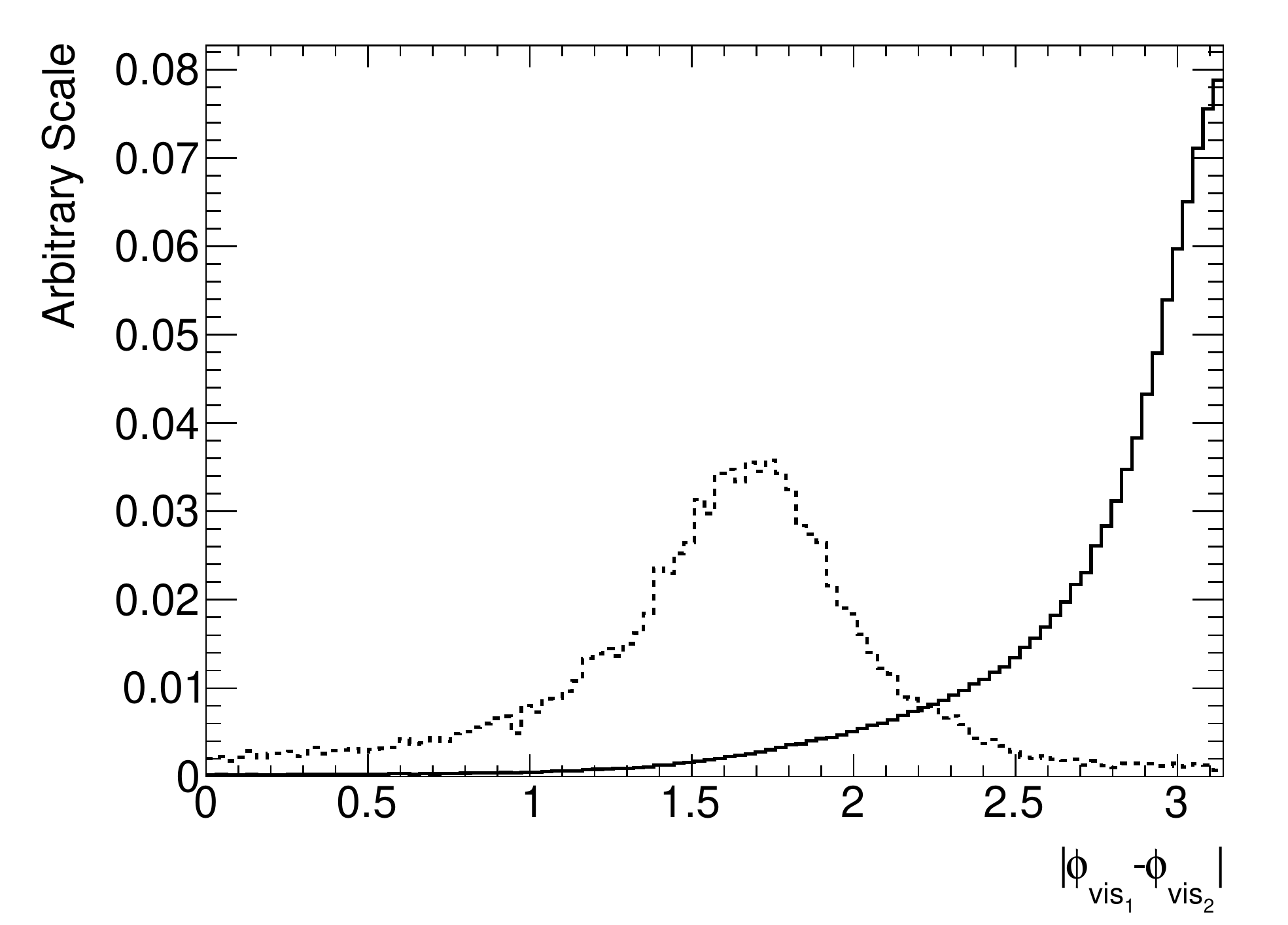}
\put(-190, 150){\textbf{(b)}}
\caption{\label{fig:dangtt_atlas} (a) The distribution of $(\theta_{\uptau_1,\uptau_2}-\theta_{vis_1,vis_2})/\theta_{vis_1,vis_2}$ after the event selection. (b) The distributions of $|\phi_{vis_1}-\phi_{vis_2}|$. In (b), the solid (dashed) histogram represents the distribution before (after) the event selection. All distributions are normalized to a unit area.} 
\end{figure}

Table~\ref{tab:ptcut} lists the thresholds on the transverse momentum while Table~\ref{tab:atlascut} lists the remaining cuts. The definitions of these variables in the tables could be found in Ref.~\cite{htt_atlas}. 
Figure~\ref{fig:dangtt_atlas}~(a) shows the distribution of $(\theta_{\uptau_1\uptau_2}-\theta_{vis_1,vis_2})/\theta_{vis_1,vis_2}$ and the relation $\theta_{\uptau_1\uptau_2}\simeq\theta_{vis_1,vis_2}$ still holds well. The distributions of $|\phi_{vis_1}-\phi_{vis_2}|$ before and after the event selection are shown in Fig.~\ref{fig:dangtt_atlas}~(b). Most of the back-to-back events are abandoned by the selections. It is then expected that the CAT would work well as $|\phi_{vis_1}-\phi_{vis_2}|$ is far from $\pi$ after the event selection.
\begin{table}[htbp]
\center{}
\caption{(color online)\label{tab:ptcut}
Summary of the transverse momentum thresholds and rapidity cuts applied in the analysis. The indices 1 and 2 denote the leading (highest $p_{T}$) and sub-leading final state objects. 
}
\begin{tabular}{l l}
\hline
Channel & Analysis level thresholds $p_{T}$ [GeV/$c$] \\
\hline
$\uptau_e\uptau_e$ & $p_{T}^{e_1}>15$, $ p_{T}^{e_2}>15$ \\
$\uptau_e\uptau_\upmu$ & $p_{T}^e>26$, $ p_{T}^\upmu>10$\\ 
$\uptau_\upmu\uptau_\upmu$ & $p_{T}^{\upmu_1}>20$, $ p_{T}^{\upmu_2}>10$ \\
$\uptau_e\uptau_h$ & $p_{T}^e>26$, $ p_{T}^{\uptau}>20$ \\
$\uptau_\upmu\uptau_h$ & $p_{T}^{\upmu}>26$, $ p_{T}^{\uptau}>20$\\
$\uptau_h\uptau_h$ & $p_{T}^{\uptau_1}>35$, $ p_{T}^{\uptau_2}>25$\\
\hline
& $|\eta(e)|<2.47$, $|\eta(\upmu)|<2.5$, $|\eta(\uptau_h)|<2.47$\\
\hline
\end{tabular}
\end{table}

\begin{table}[htbp]
\center{}
\caption{\label{tab:atlascut}
Summary of the selection criteria applied in the analysis. The definitions of the variables can be found be in Ref.~\cite{htt_atlas}.
}
\begin{tabular}{l l}
\hline
Channel & Selection cuts \\
\hline
$\uptau_l\uptau_l$ & Two opposite-sign leptons\\
& 30 GeV$<m_{\uptau\uptau}^{vis}<100 (75)$ GeV/$c^2$ for DF (SF) events \\
& $\Delta \phi_{ll}<2.5$ \\
& $E_{T}^{miss} > 20 (40)$ GeV for DF (SF) events \\
& $E_{T}^{miss,HPTO}>40$ GeV for SF events \\
& $p_{T}^{l_1}+p_{T}^{l_2}>35$ GeV/$c$ \\
& Events with a $b$-tagged jet with $p_{T}>25$ GeV$/c$ are rejected \\
& $0.1<x_{\uptau_1}, x_{\uptau_2}<1$ \\
& $m_{\uptau\uptau}^{col}>m_{\rm Z}-25$ GeV/$c^2$ \\
& At least one jet with $p_{T}>40$ GeV/$c$ \\
\hline
$\uptau_l\uptau_h$ & One lepton and one $\uptau_h$ candidate with opposite charges \\
& $m_{T}<70$ GeV/$c^2$ \\
& Events with a $b$-tagged jet with $p_{T}>30$ GeV/$c$ are rejected \\
\hline
$\uptau_h\uptau_h$ & Two opposite-sign $\uptau_h$ candidates\\
& $E_{T}^{miss}>20$ GeV \\
& $E_{T}^{miss}$ points between the two visible taus in $\phi$, or $min[\Delta\phi(\uptau,E_{T}^{miss})]<\pi/4$ \\
& $0.8<\Delta R(\uptau_{h1},\uptau_{h_2})<2.4$ \\
& $\Delta\eta(\uptau_{h1}, \uptau_{h2})<1.5$\\
\hline
All & $p_{T}^H>100$ GeV/$c$ \\
\hline
\end{tabular}
\end{table}

The performance of different mass reconstruction techniques is compared in Table~\ref{tab:performance}. Through the comparison, we find that: 1) all the three techniques recover the mass value of the Higgs boson with little bias; 2) only considering the reconstruction efficiency and the mass resolution, the technique in this work seems better than the CAT and the MMC. 3) the CAT works well and gives the similar mass resolution to our technique. The high-mass tail in the CAT disappears and the efficiency also increases.  However, the improvements are due to the specific selection criteria in Ref.~\cite{htt_atlas}. For example, for the $\uptau_l\uptau_l$ channel, it is already required that $0<x_{1}, x_2<1$ in Table~\ref{tab:atlascut} and thus the efficiency for the CAT is 100\%;   4) even considering the pileup effect, our technique is likely to have a similar performance as the MMC. Taking $\uptau_l\uptau_h$ as example, the no-pileup simulation at 8~TeV gives FWHM$/m_{peak}\simeq 24\%$ in Table~\ref{tab:performance} and the pileup simulation with $\langle\upmu\rangle=50$ at 13~TeV gives FWHM$/m_{peak}\simeq 37\%$ in Table~\ref{tab:compare}. The ATLAS experiment with $\langle\upmu\rangle \simeq 21$ at 8~TeV gives FWHM$/m_{peak}\simeq 30\%$;
5) in Table~\ref{tab:performance}, we also provide the standard deviation of the $M(\uptau\uptau)$ distribution, which can be compared with that obtained using the SVFIT method by the CMS collaboration~\cite{svfit}. It seems that our technique has better performance in the mode $\uptau_l\uptau_l$ while worse performance in the mode $\uptau_h\uptau_h$. 

\begin{table}[htbp]
\center{}
\caption{\label{tab:performance}
Performances of different mass reconstruction techniques using the ATLAS selection criteria~\cite{htt_atlas} .  $m_{peak}$, FWHM  and RMS are the peak value, the full width at half maximum, and the standard deviation of the mass distribution, respectively. The uncertainty is due to the bin width 5~GeV/$c^2$ in the mass distribution. }
\begin{tabular}{l l l l l l}
\hline
Technique & Channel & $m_{peak}$ [GeV/$c^2$] & FWHM/$m_{peak}$ & RMS (GeV/$c^2$)  & Efficiency\\
\hline
& $\uptau_l\uptau_l$ & $127.5\pm2.5$ & $(20\pm4)\%$ & 14.2 & 100\% \\
This work & $\uptau_l\uptau_h$ & $127.5\pm2.5$ & $(24\pm4)\% $ & 17.5& 100\% \\
& $\uptau_h\uptau_h$ & $127.5\pm2.5$ & $(27\pm4)\%$ & 18.8& 100\% \\
\hline
 & $\uptau_l\uptau_l$ & $132.5\pm2.5$ & $(20\pm4)\%$ & 15.5 &$100\%$ \\
CAT& $\uptau_l\uptau_h$ & $127.5\pm2.5$ & $(24\pm4)\%$ & 22.2 &$76\%$ \\
& $\uptau_h\uptau_h$ & $132.5\pm2.5$ & $(27\pm4)\%$ & 20.5& $87\%$ \\
\hline
MMC (ATLAS~\cite{htt_atlas}) & $\uptau_l\uptau_h$ & $122.3$ & $\sim30\%$ & & $\sim99\%$ \\ 
\hline
 & $\uptau_l\uptau_l$ & & & 18-24 & 100\%\\
SVFIT(CMS~\cite{svfit})& $\uptau_l\uptau_h$ &&& 14-20 & 100\%\\
& $\uptau_h\uptau_h$ &&& 13-15 & 100\%\\
\hline
\end{tabular}
\end{table}

\begin{figure}[htbp]
\center
\includegraphics[width = 0.50\textwidth]{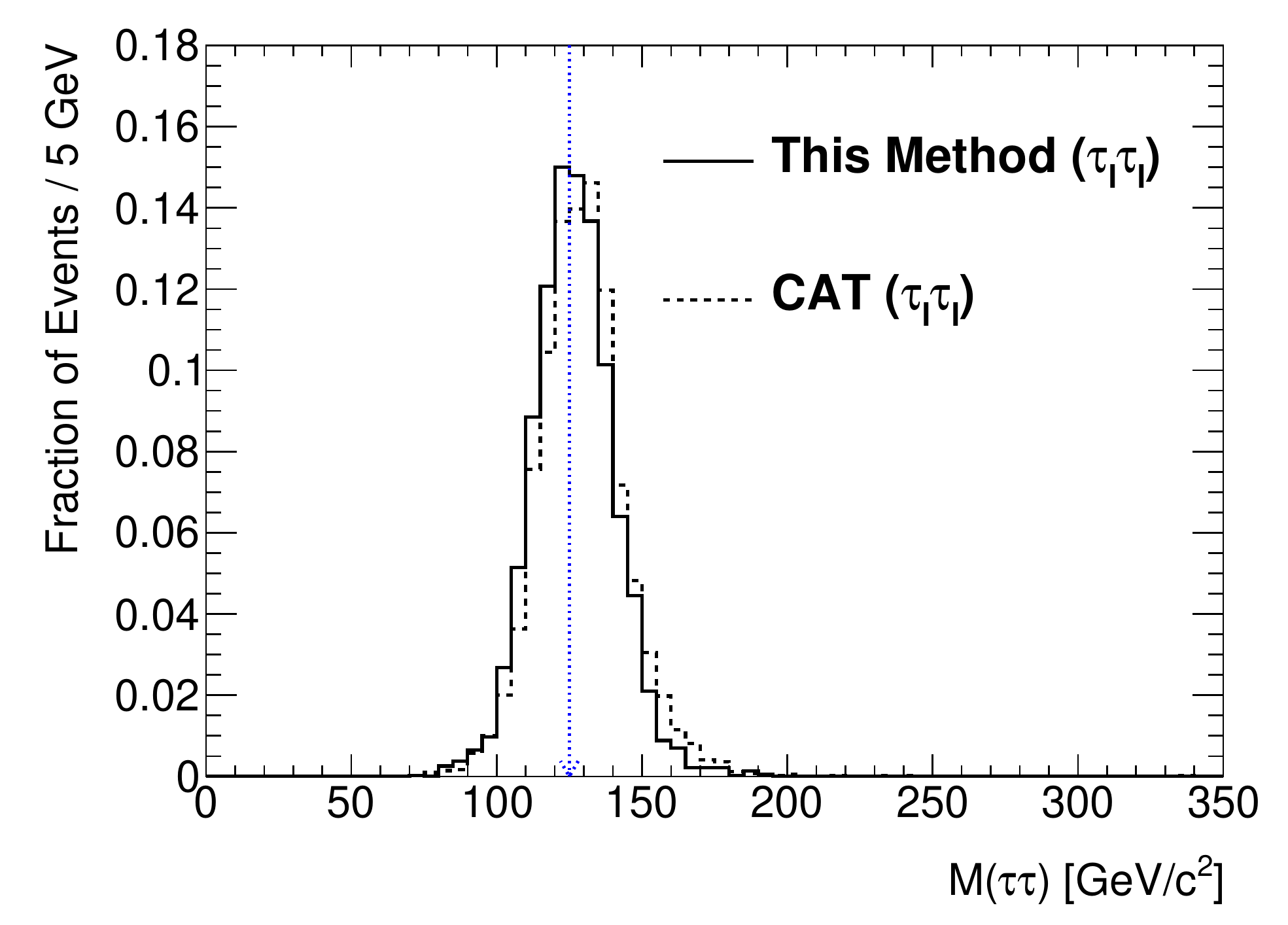}
\put(-190, 150){\textbf{(a)}}
\includegraphics[width = 0.50\textwidth]{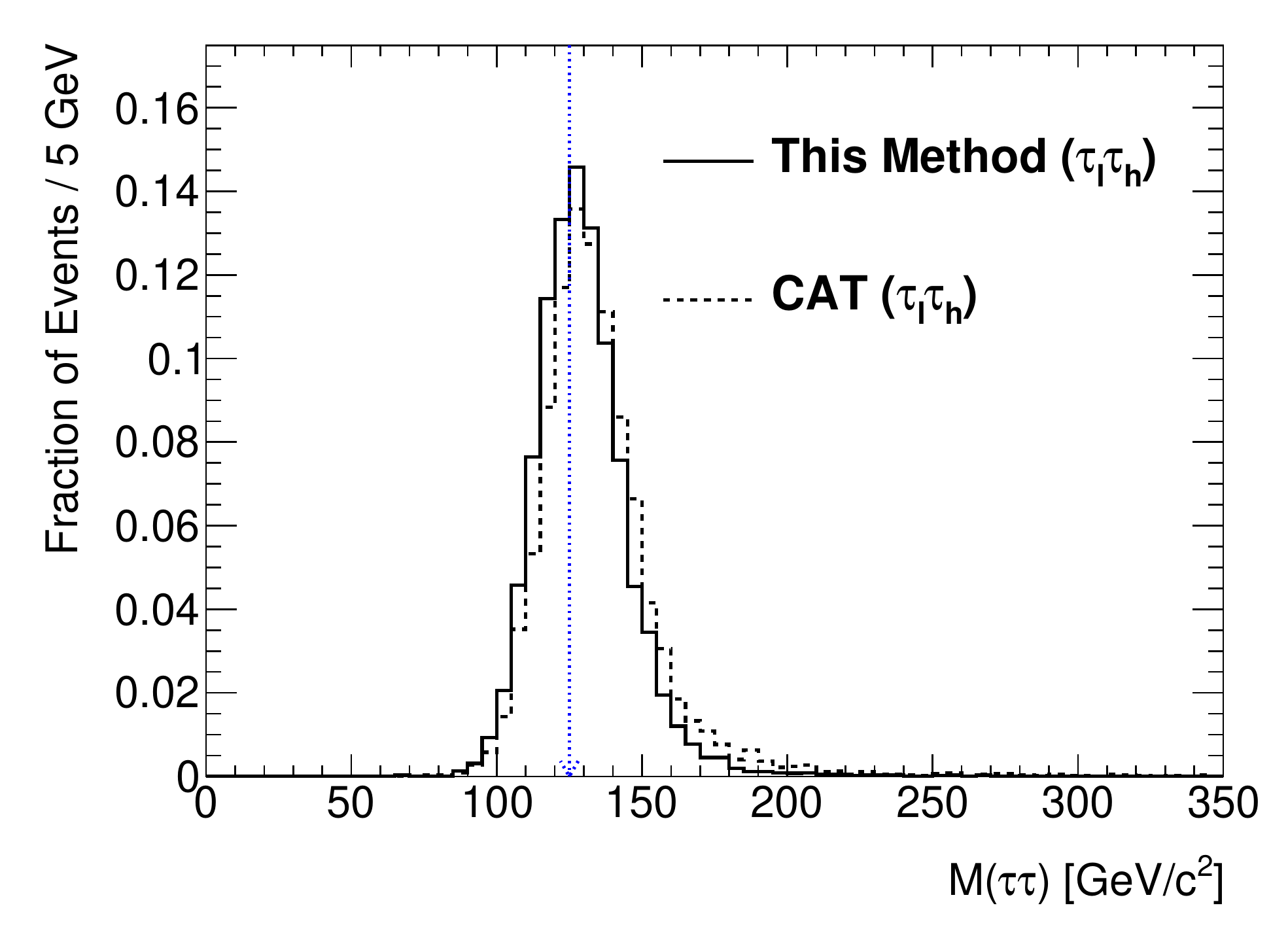}
\put(-190, 150){\textbf{(b)}}\\
\includegraphics[width = 0.50\textwidth]{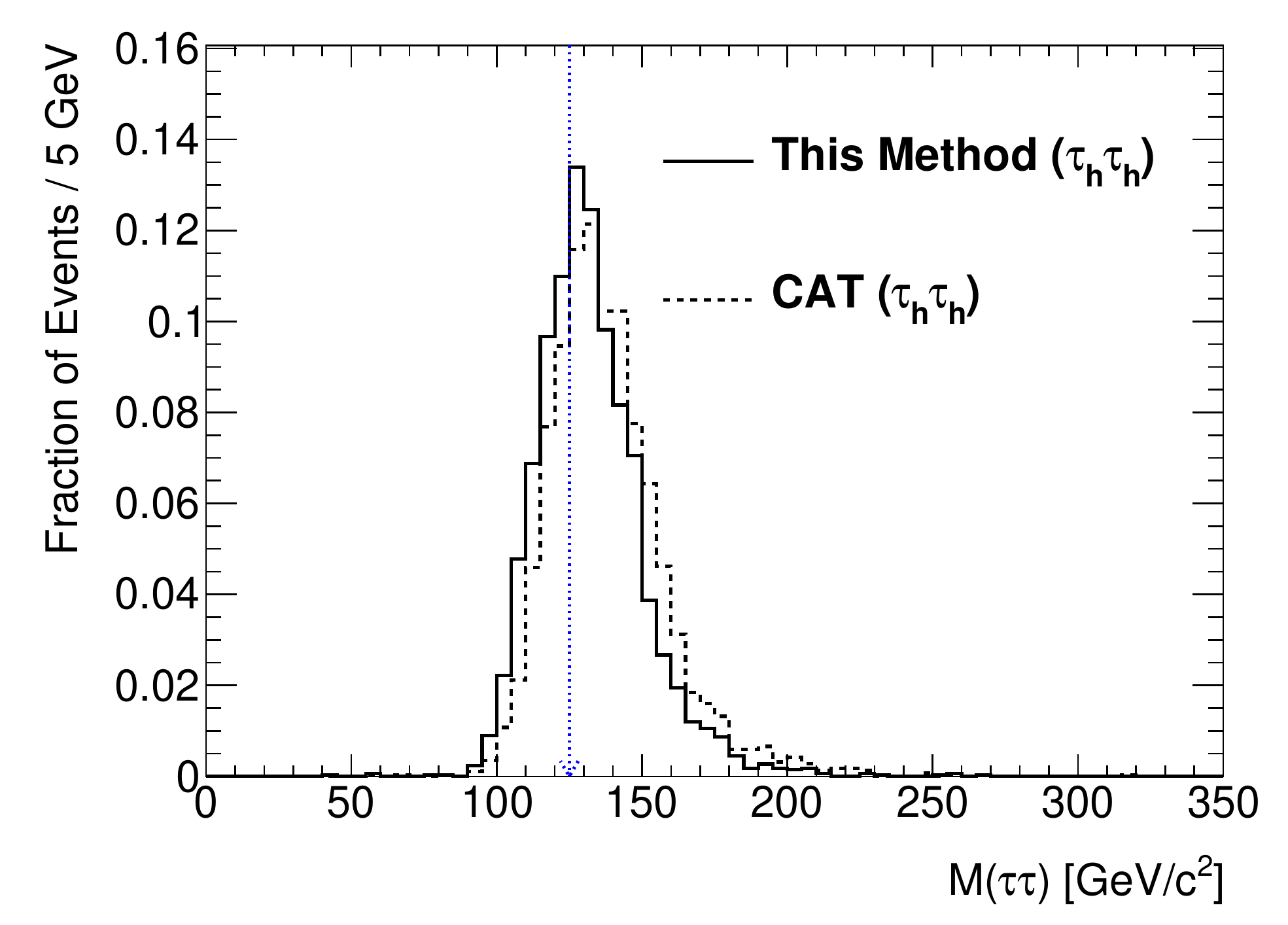}
\put(-190, 150){\textbf{(c)}}
\caption{(color online)\label{fig:hmtt_atlas}The distributions of $M(\uptau\uptau)$ for $\rm h(125)\to\uptau^+\uptau^-$ in the $\uptau_l\uptau_l$ mode (a), the $\uptau_l\uptau_h$ mode (b), and the $\uptau_h\uptau_h$ mode (c) using the ATLAS selection criteria~\cite{htt_atlas}.  The solid (dashed) histograms represent the results in this work (the collinear approximation technique). All distributions are normalized to a unit area.  The blue dashed arrow in each plot denotes the position of the generated mass.}
\end{figure}

\begin{figure}[htbp]
\center
\includegraphics[width = 0.50\textwidth]{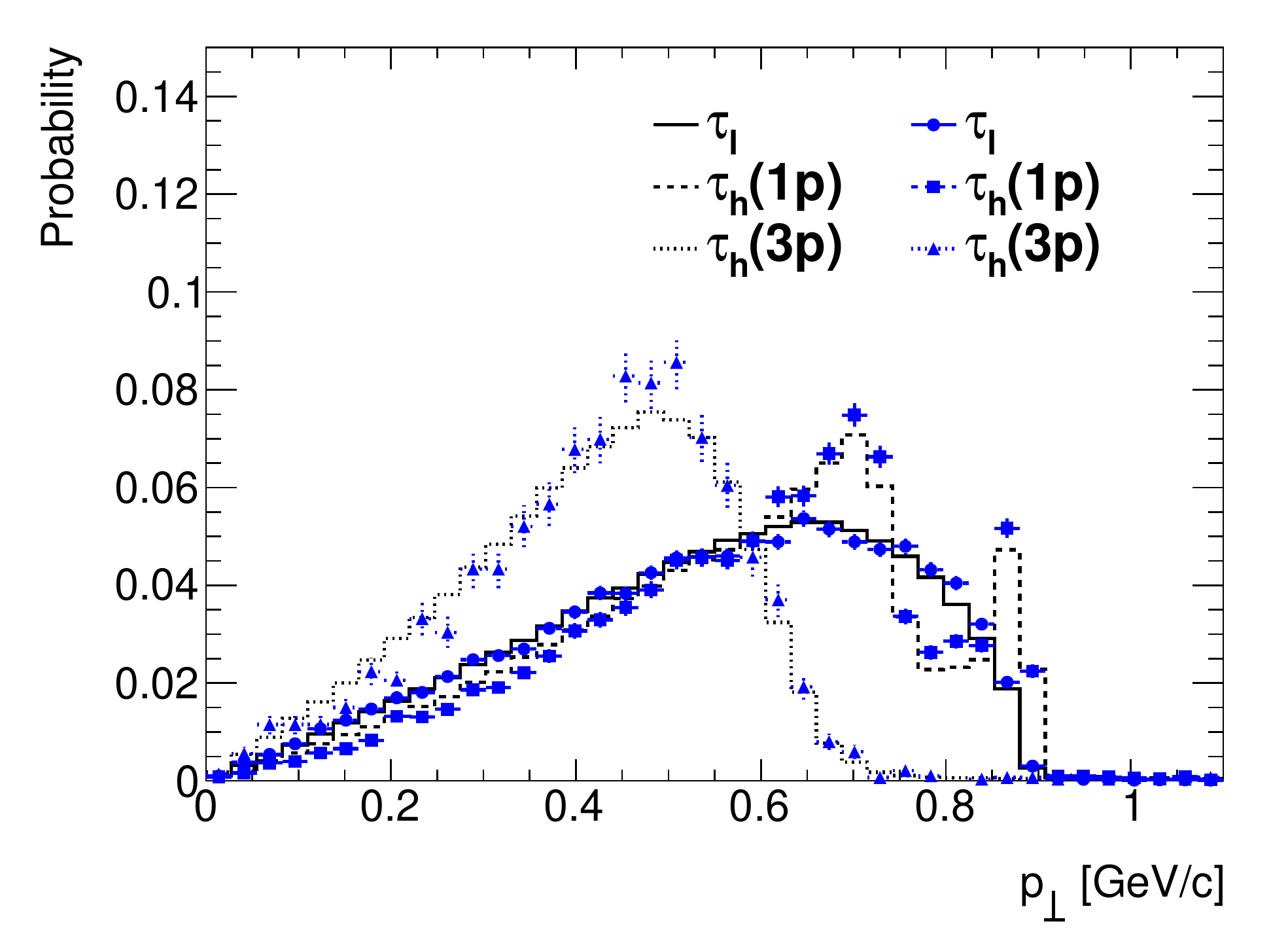}
\put(-190, 150){\textbf{(a)}}
\includegraphics[width = 0.50\textwidth]{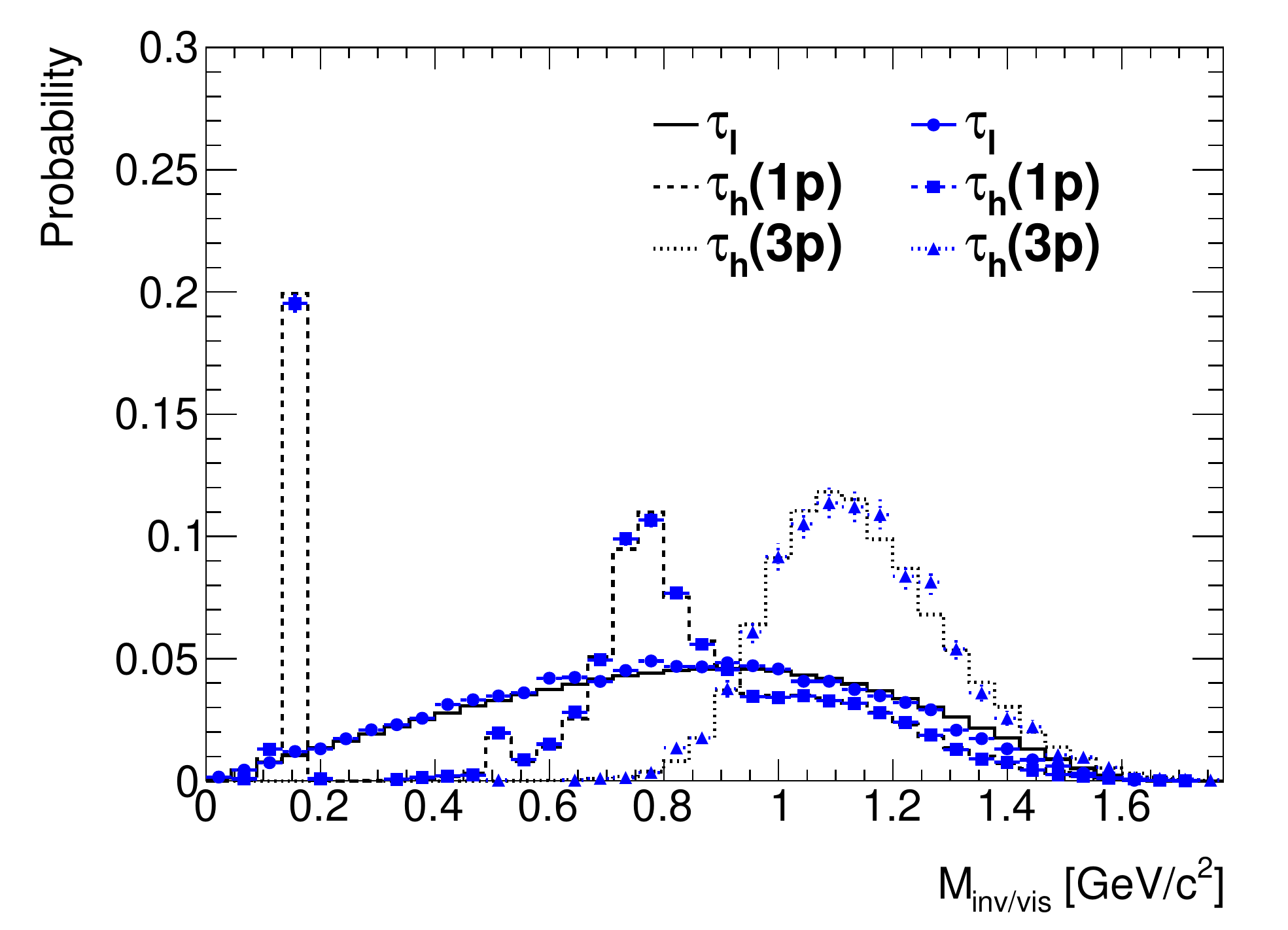}
\put(-190, 150){\textbf{(b)}}
\caption{(color online)\label{fig:invp_Mvis_ba} (a) The distributions of $p_\perp$ in the lepton mode (solid line) and the hadron mode with one charged track (dashed line) or three charged tracks (dotted line). (b) The mass distributions of the invisible neutrinos in the lepton mode (solid line) and the visible hadrons in the hadron mode with one charged track (dashed line) or three charged tracks (dotted line). The black histograms (blue markers with error bar) are the distributions before (after) the event selection. All distributions are normalized to a unit area. } 
\end{figure}

In the end of this section,  Fig.~\ref{fig:invp_Mvis_ba} compares the probability distributions $\mathcal{P}(p_\perp, m_{vis/inv})$ before and after the event selection summarized in Table~\ref{tab:ptcut} and~\ref{tab:atlascut}. It is found that the probability distributions  are not sensitive to the event selection. This point indicates that we do not have to tune the JPDs once the selection requirements are changed, and the systematic uncertainty of the shape of the reconstructed mass reconstruction due to the used JPDs may be small in this technique. 

\section{Conclusions}\label{sec:conclusion}
In summary, we propose a new technique to reconstruct the mass of a heavy resonance decaying to $\uptau^+\uptau^-$. It is based on the collinear approximation $\theta_{\uptau_1\uptau_2}\simeq\theta_{vis_1,vis_2}$. The mass $M(\uptau\uptau)$ can be determined with given $(p_{\perp}, m_{vis/inv})$ for the $\uptau$ leptons.  Through sampling $(p_{\perp}, m_{vis/inv})$ according to their joint probability distributions, the reconstructed mass is assumed to correspond to the value with the maximal probability. This method utilizes the fact that the quantities $p_{\perp}$ and $m_{{vis/inv}}$ are invariant under the boost in the direction of the $\uptau$ lepton. Based on the MC simulations of $\rm pp\to Z/h(125)/h(750)+X\to\uptau^+\uptau^-+X$, this method gives a relative mass resolution, FHWM/$m_{peak} \simeq$ 20\%-40\% using the information of missing energy with no efficiency loss. In the end, we would like to comment that this method would work better in the  Circular Electron Positron Collider (CEPC), since the well-determined initial four-momenta of the colliding beams provide more kinematic constraints.

\section{Acknowledgements}
Li-Gang Xia would like to thank Fang Dai for many helpful discussions. This work is supported by the General Financial Grant from the China Postdoctoral Science Foundation (Grant No. 2015M581062).

\vspace{-1mm}
\centerline{\rule{80mm}{0.1pt}}
\vspace{2mm}

\begin{multicols}{2}

\end{multicols}

\clearpage
\end{CJK*}
\end{document}